\documentclass[iop]{emulateapj}
\usepackage{natbib}
\usepackage{longtable}
\usepackage{amsmath}
\usepackage{epsfig}
\usepackage{float}
\usepackage{textcomp}
\usepackage{color}
\usepackage{apjfonts}
\usepackage{xspace}

\newcommand{\water}{\ensuremath{\textrm{H}_2\textrm{O}}\xspace}

\slugcomment{20 April 2017; Published in ApJ}

\shorttitle{Young Stellar Objects in W43 and W51}
\shortauthors{Saral et al.}

\begin{document}

\title{Young Stellar Objects in the Massive Star-Forming Regions W51 and W43}

\author{
G. Saral\altaffilmark{1,2,3},
J. L. Hora \altaffilmark{2},
M. Audard\altaffilmark{1},
X. P. Koenig\altaffilmark{4},
J. R. Mart\'{\i}nez-Galarza\altaffilmark{2},
F. Motte\altaffilmark{5,6},
Q. Nguyen-Luong\altaffilmark{7,8}
A. T. Saygac\altaffilmark{9,10},
H. A. Smith\altaffilmark{2}
}
\altaffiltext{1}{Department of Astronomy, University of Geneva, Ch. d'Ecogia 16, 1290 Versoix, Switzerland}
\altaffiltext{2}{Harvard-Smithsonian Center for Astrophysics, 60 Garden Street, Cambridge, MA 02138, USA}
\altaffiltext{3}{Istanbul University, Graduate School of Science and Engineering, Bozdogan Kemeri Cad. 8, Vezneciler, Istanbul, Turkey}
\altaffiltext{4}{Yale University, Department of Astronomy, 208101, New Haven, CT 06520-8101, USA}
\altaffiltext{5}{Institut de Planétologie et d'Astrophysique de Grenoble, Univ. Grenoble Alpes - CNRS-INSU, BP 53, F-38041 Grenoble Cedex 9, France}
\altaffiltext{6}{AIM Paris-Saclay, CEA/IRFU - CNRS/INSU - Univ. Paris Diderot, Service d'Astrophysique, CEA-Saclay, F-91191 Gif-sur-Yvette Cedex, France}
\altaffiltext{7}{National Astronomical Observatory of Japan, Chile Observatory, 2-21-1 Osawa, Mitaka, Tokyo 181-8588, Japan}
\altaffiltext{8}{Korea Astronomy and Space Science Institute, 776 Daedeok daero, Yuseoung, Daejeon 34055, Korea}
\altaffiltext{9}{Istanbul University, Faculty of Science, Astronomy and Space Sciences Department, Istanbul-Turkey}
\altaffiltext{10}{Istanbul University Observatory Application and Research Center, Vezneciler, Fatih, Istanbul, Turkey}
    
\begin{abstract}
We present the results of our investigation of the star-forming complexes W51 and W43, two of the brightest in the first Galactic quadrant. In order to determine the young stellar object (YSO) populations in W51 and W43 we used color-magnitude relations based on \textit{Spitzer} mid-infrared and 2MASS/UKIDSS near-infrared data. We identified 302 Class I YSOs and 1178 Class II/transition disk candidates in W51, and 917 Class I YSOs and 5187 Class II/transition disk candidates in W43. We also identified tens of groups of YSOs in both regions using the Minimal Spanning Tree (MST) method. We found similar cluster densities in both regions, even though Spitzer was not able to probe the densest part of W43. By using the Class II/I ratios, we traced the relative ages within the regions, and based on the morphology of the clusters, we argue that several sites of star formation are independent of one another in terms of their ages and physical conditions. We used spectral energy distribution-fitting to identify the massive YSO (MYSO) candidates since they play a vital role in the star formation process, and then examined them to see if they are related to any massive star formation tracers such as UC\ion{H}{2} regions, masers, or dense fragments. We identified 17 MYSO candidates in W51, and 14 in W43, respectively, and found that groups of YSOs hosting MYSO candidates are positionally associated with \ion{H}{2} regions in W51, though we do not see any MYSO candidates associated with previously identified massive dense fragments in W43.
 \end{abstract}


\section{Introduction} \label{sec:int}

Near-infrared and mid-infrared observations have revealed a broad range of stellar densities of YSOs in star-forming regions \citep{bre10}. While we have a detailed picture of the formation of low-mass stars with advances in observational techniques and improved theoretical knowledge in the last decade, our understanding of massive star formation is still lagging. Massive stars form and evolve quickly and they produce strong winds and outflows which heat, ionize, and disrupt their natal molecular cloud, which makes the observations of early evolutionary stages difficult. In addition, massive stars are forming in a clustered mode by interacting with each other as seen in Orion \citep{lad92,eva09,meg16} or Cygnus-X \citep{mot07,bee10,kry14,sch16}. In order to understand the star formation process, it is necessary to understand that the formation of massive clusters and the Galactic star-forming regions are the ideal places for this purpose, since they host many active sites of star formation in different environments at various stages of evolution.

We started a detailed study of a sample of mini-starburst complexes \citep{mot03} to better understand how massive stars form and affect their environments and the effects that they have on the formation of lower-mass stars \citep{sar13}. Precursor clouds of young massive clusters (YMC) \citep{bre12,gin12,wal16} are thought to be the potential descendant of mini-starburst clouds/clumps \citep{mot03,2011A&A...529A..41N,2011A&A...535A..76N,lou14,ngu16}. YMCs (M $\gtrsim$ $10^4$ M$_\odot$, ages $\lesssim$ 100 Myr) are stellar systems that host a very large number of YSOs with a range of masses including massive ones \citep{lon14}. The first paper of this study \citep{sar15} focused on W49, which is considered to be a mini-starburst region hosting a YMC precursor. We identified thousands of YSOs and tens of massive YSO (MYSO) candidates in W49, which shows a compact massive cluster in its center.

In this second paper, we investigate two YMC precursors, W51 \citep{wal16} and W43, which are known to host two mini-starburst ridges \citep{ngu13}. They are both among the brightest and most massive of the first Galactic quadrant. We present an imaging and photometric analysis of W51 and W43 with deep IR data from 1 to 24 $\mu$m in order to investigate the MYSOs and the embedded clusters forming around them. 

W51 was first detected as an \ion{H}{2} region by \citet{wes58} and later identified as a molecular cloud by CO emission \citep{pen71}. It hosts the star forming regions W51A and W51B and a supernova remnant W51C, with a total mass of $\sim1.2\times10^{6}$ \citep{car98} and a stellar mass of $\sim10^{4}$ \citep{kum04}. \citet{xu09} used methanol masers to measure the trigonometric parallaxes and found a distance of $\displaystyle{5.1^{+2.9}_{-1.4}}$ kpc. \citet{sat10} reported the distance as $\displaystyle{5.4^{+0.31}_{-0.28}}$ kpc with higher accuracy by using \water masers and the trigonometric parallax method. There are studies based on maser emissions and velocity measurements showing the interaction between the star-forming region W51B and the supernova remnant W51C \citep[i.e.,][]{gre97,gin15}.  \citet{kan09} studied the YSO population of W51 by using \textit{Spitzer} and 2MASS data and combining both the color selection criteria from \citet{sim07} and the spectral energy distribution (SED) slope classification \citep{lad91,and93,gre94}, in addition to SED models from \citet{rob06}. This resulted in the identification of 737 YSO candidates. Although there is a slight difference between the cataloged region in \citet{kan09} study and in this paper, they have 91\% of the sources in common; and among them, 54\% of the sources are YSO candidates in both studies. We compare our results with this earlier work in Section \ref{sec:iraccolor}.

W43 is called a mini-starburst \citep{mot03} with several tens of cores forming high-mass stars, among which are one UC\ion{H}{2} region, a handful of maser sources, and at least three very massive Class 0s with $\sim10^4$ ${\rm L}_{\sun}$ in the massive cloud. The starburst cluster was first detected as an unresolved source by \citet{les85} in near-IR images and then confirmed as a massive cluster (W43-Main) by \citet{blu97}. The complex hosts two bright clouds: W43-Main (G30.8+0.02), and W43-South (G29.96-0.02). The detailed analysis of molecular and atomic gas tracers by \citet{2011A&A...529A..41N} showed that these two regions are connected and the complex extends over 140 pc at $l$ = (29$^{\circ}$$-$32$^{\circ}$), $b$ = (-1$^{\circ}$ $-$ +1$^{\circ}$). The two clouds are surrounded by an atomic gas envelope with a diameter of 290 pc. The connection between W43-Main and W43-South has been confirmed by \citet{car13}, and \citet{mot14} confirmed the presence of envelope, giving its diameter as 270 pc and a mass of $M_\mathrm{HI,env}\sim3\times10^6$ M$_\odot$. On the other hand, recently,  \citet{bia17} argued that the large HI gas can be explained by several HI to H2 transition layers along the sightlines instead of one large HI envelope based on 21 cm observations and a theoretical model for the HI-to-H2 transition.\citet{2011A&A...529A..41N} calculated its kinematic distance as 6~kpc and placed it at the meeting point of the Galactic bar and the Scutum-Centaurus Arm, while \citet{zha14} redetermined the distance as $\displaystyle{5.49^{+0.39}_{-0.34}}$~kpc using the trigonometric parallax method by studying the masers G029.86-00.04, G029.95-00.01, G031.28+00.06 and G031.58+00.07. 

In Section \ref{sec:obs}, we describe the observations, data reduction techniques, and our near- and mid-IR source catalog and YSO classification. In Section \ref{sec:clustering}, we present the clustering analysis. Section \ref{sec:sed} contains the SED fitting results for massive YSO candidates. In Section \ref{sec:discuss} we present the massive star formation tracers, we discuss the star formation history and compare W51 and W43 to other star-forming regions. Finally, in Section \ref{sec:sum}, we summarize our results.

\section{Observations and Methods} \label{sec:obs}

\subsection{IRAC Imaging} \label{iracimaging}

\begin{figure*}
\centering
\includegraphics[width=15cm]{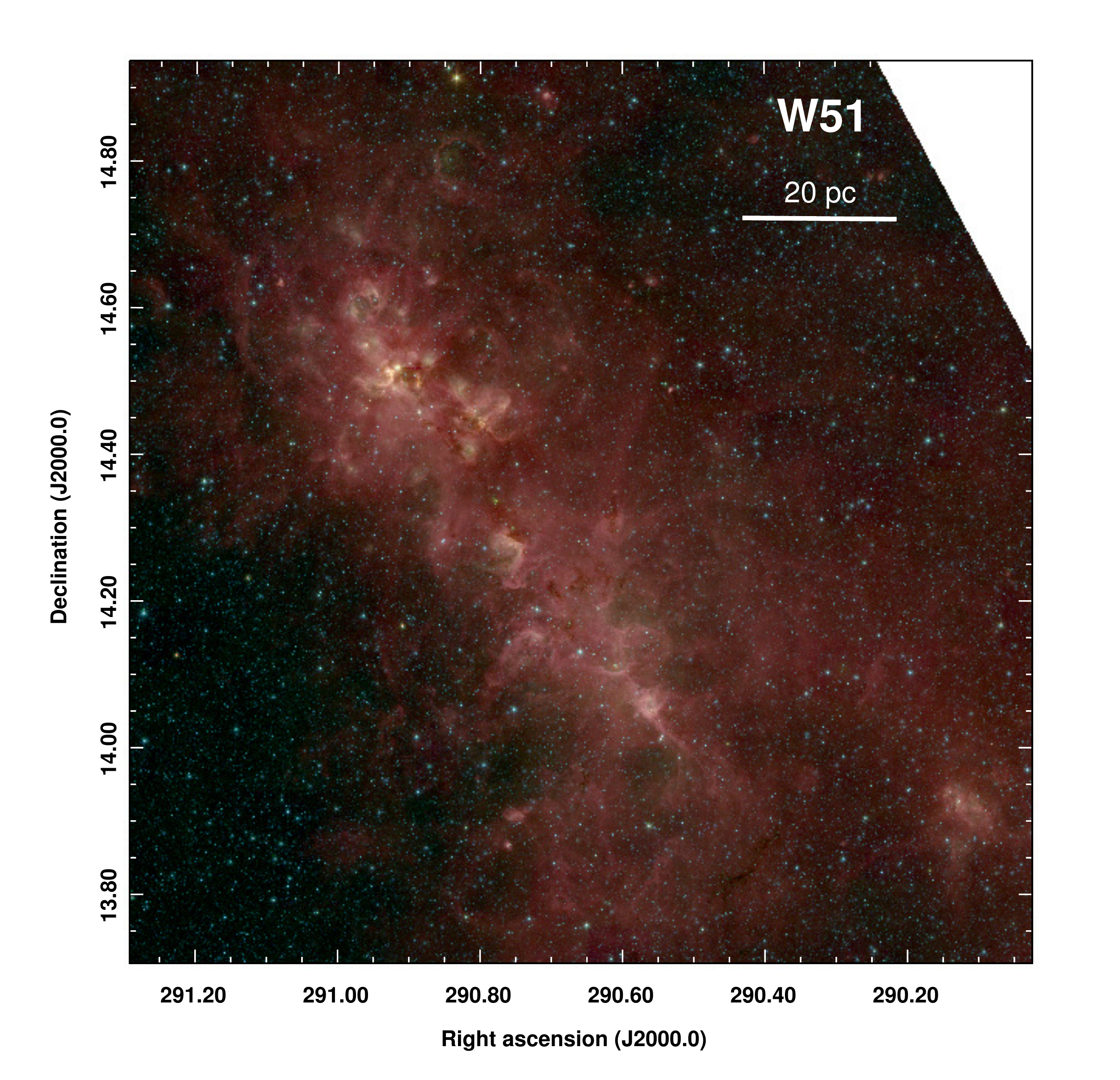}
\caption{The entire W51 is shown in Spitzer IRAC bands (blue: 3.6 $\mu$m, green: 4.5 $\mu$m, red: 8.0 $\mu$m).}\label{fig:W51GMC}
\end{figure*}

We used the mid-infrared \textit{Spitzer} data from several programs obtained with the \textit{Spitzer} IRAC instrument \citep{faz04} at $3.6$, $4.5$, $5.8$, and $8.0$~$\mu$m. The 2 s frame time data taken at all wavelengths in 2007 and prior years are from the GLIMPSE Legacy Survey \citep[][program ID 187]{chu09}. The data obtained in 2012 and afterward were performed during the warm mission, so only $3.6$ and $4.5$~$\mu$m data were obtained. We also used 30 s frame time data from other programs taken in October 2004, 2005, and June 2013, with the following program IDs: 2313 (PI: M. Kuchner), 20026 (PI: T. von Hippel), and 90095 (PI: K. Luhman). We also obtained 30 s data in High Dynamic Range (HDR) mode in December 2014 (project ID: 10012, PI: J. Hora), which performs consecutive individual observations with frame times of 1.2 s and 30 s. An RGB image of the W51 region constructed from the 2 s data is shown in Figure~\ref{fig:W51GMC}. 

\begin{figure*}
\centering
\includegraphics[width=15cm]{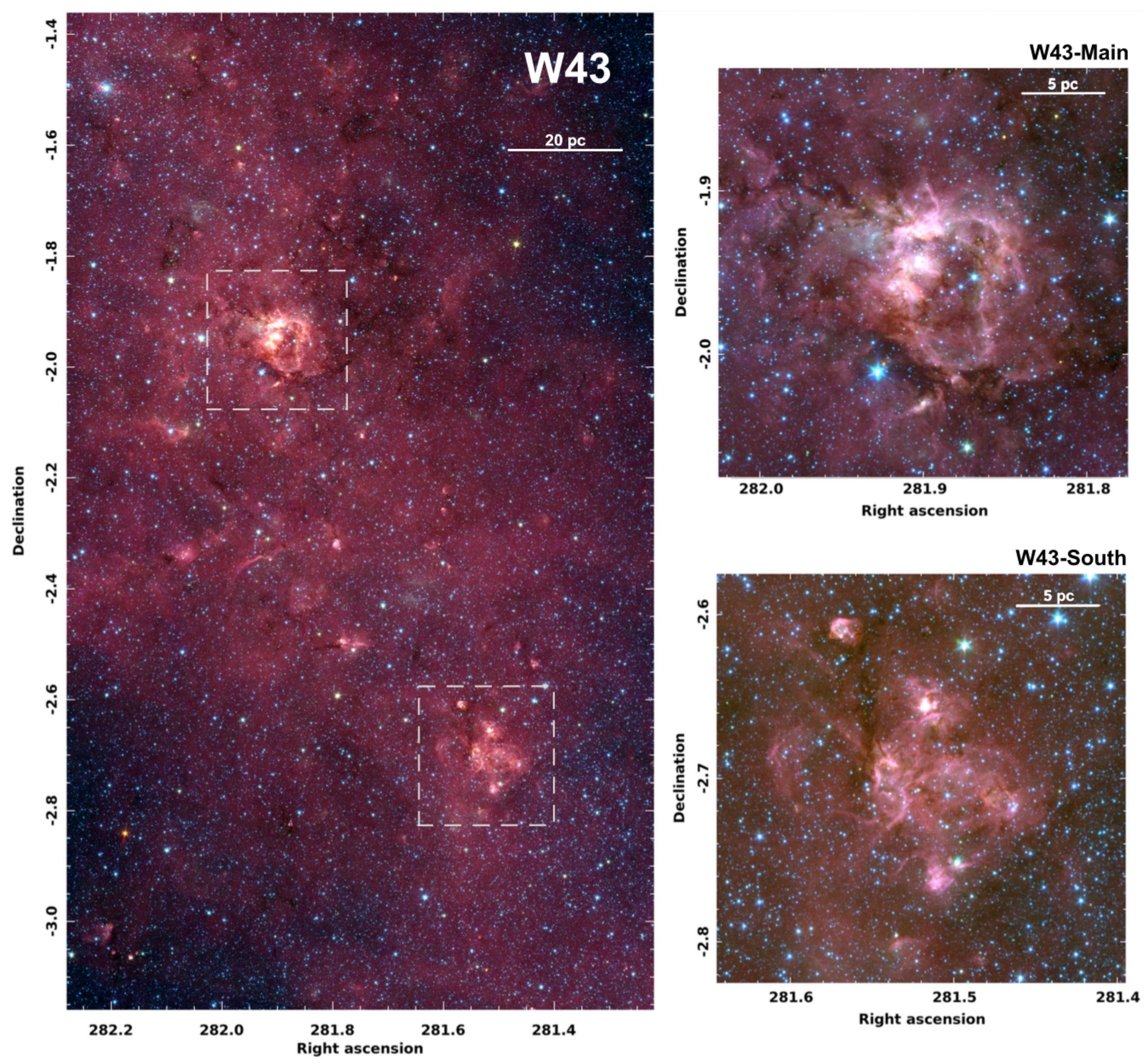}
\caption{Left: the entire W43 is shown in Spitzer IRAC bands. W43-Main and W43-South are shown within dashed boxes. Right: W43-Main is shown in the upper right panel and W43-South is shown in the lower right panel (blue: 3.6 $\mu$m, green: 4.5 $\mu$m, red: 8.0 $\mu$m).}\label{fig:W43GMC}
\end{figure*}

The IRAC data for the W43 region are from the GLIMPSE Legacy Survey \citep[][program IDs 186, 30570]{ben03,chu09} and 30 s HDR data were taken with program IDs 10012 and 80058 (PI: J. Hora). An RGB image of the W43 is shown in Figure~\ref{fig:W43GMC}. The large image was constructed from the 2 s frames; the smaller images of W43-Main and W43-South were constructed from the 30 s HDR frames for the $3.6$ and $4.5$~$\mu$m wavelengths and the 2 s images at 8.0~$\mu$m. We list the dates and coordinates of each Astronomical Observation Request (AOR) in Tables~\ref{W51aors} and ~\ref{W43aors}. 

We generated the mosaic images from the standard basic calibrated data (BCD) products (processed with the Spitzer IRAC pipeline version S18.25.0 and S19.1.0).  Automated source detection and aperture photometry were carried out using PhotVis 1.10 \citep{gut04,gut08}. PhotVis utilizes a modified DAOphot \citep{ste87} source-finding algorithm. Aperture photometry was performed with an aperture of 2.4$\arcsec$ radius and using a background annulus of inner and outer radii 2.4$\arcsec$, and 7.2$\arcsec$ respectively. The IRAC PSF is 1.66$\arcsec$, 1.72$\arcsec$, 1.88$\arcsec$, and 1.98$\arcsec$ at $3.6$, $4.5$, $5.8$ and $8.0$~$\mu$m, respectively \citep{faz04}, corresponding to 0.044$-$0.052 pc for W51 (5.4 kpc) and 0.045$-$0.054 pc for W43 (5.6 kpc). In the first paper we had estimated the completeness magnitudes in the W49 region for $3.6$, $4.5$, $5.8$, and $8.0$~$\mu$m as 15.1, 14.7, 12.35, and 12.12, respectively, and since the exposure times are same and the distances are similar, we assume the completeness estimate will be similar for W51 and W43. 

\begin{deluxetable}{llllcl}
\tabletypesize{\scriptsize}
\tablecaption{Astronomical Observation Requests for W51\label{W51aors}}
\tablehead{\colhead{AORKEY} & \colhead{Date} & \colhead{R.A.} & \colhead{Decl.} & \colhead{Frame } & \colhead{IRAC}
\\ \colhead{ } & \colhead{(UT)} & \colhead{(J2000)} & \colhead{(J2000)} & \colhead{Time} & \colhead{Reduction}
\\ \colhead{ } & \colhead{ } & \colhead{ } & \colhead{} & \colhead{ } & \colhead{Pipeline ver.}}
\startdata 
12241920 & 2004 Oct 10 & 19:21:33 & 14:20:08 & 2 & S18.25.0 \\
12242176 & 2004 Oct 10 & 19:20:49 & 14:00:17 & 2 & S18.25.0 \\
12243200 & 2004 Oct 10 & 19:19:22 & 13:20:24 & 2 & S18.25.0 \\
12243456 & 2004 Oct 10 & 19:20:05 & 13:40:20 & 2 & S18.25.0 \\
12244480 & 2004 Oct 10 & 19:22:17 & 14:40:03 & 2 & S18.25.0 \\
12244736 & 2004 Oct 11 & 19:25:15 & 15:59:30 & 2 & S18.25.0 \\
10127616 & 2004 Oct 29 & 19:21:40 & 14:04:36 & 30 & S18.25.0 \\
13828864 & 2005 Oct 21 & 19:21:40 & 14:40:36 & 30 & S18.25.0 \\
48640768 & 2013 Jun 26 & 19:21:40 & 14:40:36 & 30 & S19.1.0 \\
51112704 & 2014 Dec 04 & 19:21:45 & 13:49:21 & 30\tablenotemark{*}& S19.1.0 \\
51112448 & 2014 Dec 10 & 19:19:51 & 14:01:37 & 30\tablenotemark{*} & S19.1.0 \\
\enddata
\tablenotetext{*}{HDR data}
\end{deluxetable}

\begin{deluxetable}{llllcl}
\tabletypesize{\scriptsize}
\tablecaption{Astronomical Observation Requests for W43\label{W43aors}}
\tablehead{\colhead{AORKEY} & \colhead{Date} & \colhead{R.A.} & \colhead{Decl.} & \colhead{Frame } & \colhead{IRAC}
\\ \colhead{ } & \colhead{(UT)} & \colhead{(J2000)} & \colhead{(J2000)} & \colhead{Time} & \colhead{Reduction}
\\ \colhead{ } & \colhead{ } & \colhead{ } & \colhead{} & \colhead{ } & \colhead{Pipeline ver.}}
\startdata 
9237248 & 2004 Apr 22 & 18:49:12 & -01:06:39 & 2 & S18.25.0\\
9241344 & 2004 Apr 22 & 18:43:55 & -03:40:47 & 2 & S18.25.0\\
9242880 & 2004 Apr 22 & 18:45:22 & -02:59:36 & 2 & S18.25.0\\
9244416 & 2004 Apr 22 & 18:44:40 & -03:18:54 & 2 & S18.25.0\\
9244672 & 2004 Apr 22 & 18:45:55 & -02:43:12 & 2 & S18.25.0\\
9240576 & 2004 Apr 22 & 18:47:43 & -01:50:48 & 2 & S18.25.0\\
9246208 & 2004 Apr 22 & 18:47:02 & -02:10:14 & 2 & S18.25.0\\
9246464 & 2004 Apr 22 & 18:48:28 & -01:28:42 & 2 & S18.25.0\\
21269248 & 2007 May 12 & 18:39:17 & -01:32:16 & 2 & S18.25.0 \\
21272320 & 2007 May13 & 18:53:31 & -02:28:02 & 2 & S18.25.0 \\
21326080 & 2007 May 13 & 18:54:17 & -02:44:47 & 2 & S18.25.0\\
21274624 & 2007/ May 13 & 18:54:13 & -03:01:57 & 2 & S18.25.0\\
45847296 & 2012 Nov 24 & 18:47:55 & -01:35:37 & 30\tablenotemark{*} & S19.1.0\\
45847552 & 2012 Nov 24 & 18:46:23 & -02:31:00 & 30\tablenotemark{*} & S19.1.0\\
45847808 & 2012 Nov 25 & 18:48:55 & -02:23:00 & 30\tablenotemark{*} & S19.1.0\\
45848064 & 2012 Nov 21 & 18:56:24 & -01:48:00 & 30\tablenotemark{*} & S19.1.0\\
45848320 & 2012 Nov 25 & 18:50:09 & -01:27:54 & 30\tablenotemark{*} & S19.1.0\\
51107840 & 2014 Nov 29 & 18:46:52 & -01:51:23 & 30\tablenotemark{*} & S19.1.0\\
51108352 & 2014 Jun 19 & 18:47:43 & -01:48:24 & 30\tablenotemark{*} & S19.1.0\\
51108608 & 2014 Nov 29 & 18:48:04 & -02:16:50 & 30\tablenotemark{*} & S19.1.0\\
51108864 & 2014 Nov 29 & 18:48:19 & -01:46:17 & 30\tablenotemark{*} & S19.1.0\\
\enddata
\tablenotetext{*}{HDR data}
\end{deluxetable}

\subsection{Source Catalog} \label{sec:cat}

In the W43 region, within an area of size $\triangle$\emph{$\alpha$} $\times$ $\triangle$\emph{$\delta$} = 2\fdg0 $\times$ 2\fdg3, centered at (\emph{$\alpha$}, \emph{$\delta$}) = (281\fdg8, -2\fdg2), 641,732 sources were detected in the IRAC images. Among these, 118,365 sources have photometry in all four IRAC filters. In W51, within an area of size $\triangle$\emph{$\alpha$} $\times$ $\triangle$\emph{$\delta$} = 1\fdg5 $\times$ 1\fdg5, centered at (\emph{$\alpha$}, \emph{$\delta$}) = (290\fdg6, 14\fdg2), 285,252 sources were detected. Among these, 37,932 sources have photometry in all four IRAC filters.

After performing the photometry in each of the four IRAC bands, we combined the catalogs of short (2 s) and long (30 s) frames using a band-merging process that takes all of the sources from the short frames that are brighter than a certain cutoff (magnitude 11, 10.4, 8.7, and 8.2 for $3.6$, $4.5$, $5.8$, and $8.0$~$\mu$m, respectively) based on the saturation limit of 30 s frames. All of the sources detected by \textit{Spitzer} were matched to Two Micron All Sky Survey (2MASS) Point Source Catalog \citep{skr06} sources by the PhotVis photometry routine. We also cross-matched the IRAC Catalog with the UKIRT Infrared Deep Sky Survey (UKIDSS)\footnote[5]{UKIDSS uses the UKIRT Wide Field Camera \citep{cas07} on the United Kingdom Infrared Telescope, and the UKIDSS project is defined by \citet{law07} Galactic Plane Survey \citep{luc08} data from Data Release 8 (DR8PLUS), which are deeper and have better spatial resolution than those of 2MASS.} We used a maximum of 1$\arcsec$ for the matching radius to allow a reasonable error in source positions and also to prevent false matches. We took into account the saturation limit in the UKIDSS survey which is near 12.65, 12.5, and 12 mag in $J$, $H$, and $K_{s}$, respectively, and used 2MASS photometry when the source was saturated in UKIDSS (20\% of the data in UKIDSS $JHK$ photometry was saturated in W51 and 25\% in W43). The UKIDSS data have small but measurable zero-point photometric offsets from 2MASS, therefore we calculated the mean and standard deviations of the magnitude residuals between 2MASS and UKIDSS in both W43 and W51 catalogs and applied a mean offset to UKIDSS data of 0.06, $-$0.06, and $-$0.01 mag to the $J$, $H$, and $K_{s}$ bands, respectively. In addition, due to very small photometric errors in the UKIDSS catalog (e.g., $<$0.001 mag for 13 mag sources), we adjusted the values by adding 0.02 mag in quadrature, which imposes an error floor of 0.02 mag but does not affect the larger errors. Lastly, we matched our catalog with the MIPSGAL Archive 24~$\mu$m data \citep{gut15}. We used 1$\arcsec$ as a maximum radial tolerance based on Monte Carlo simulations that give a mean false match probability of 6~\% and 5.5~\% for W51, and W43, respectively.

In the case of multiple matches, we selected the closest IRAC object to the MIPS source. In W51, only 15 MIPS sources had a second IRAC object within the 1$\arcsec$ search radius that could possibly contribute to its 24~$\mu$m flux. However, these sources were either unclassified or classified as photospheres and only one source was classified as a YSO candidate, while the second IRAC object was an unclassified object. Similarly, in W43, 24 sources had a second IRAC object within the 1$\arcsec$ search radius. Within these matches, in only one pair were two IRAC sources were identified as YSO candidates that might be contributing to each other's 24~$\mu$m flux. However, in the rest of the 11 pairs, we see that the second IRAC source is unclassified. As a summary, there are only two YSO candidates in W43 that might be contributing to each other's 24~$\mu$m flux;  we consider this as having a minimal effect on our analysis.

A summary of the source catalog and source-matching results can be found in Table~\ref{datasummary} and in Table~\ref{datasourcematch}, respectively. In the end, we generated  a final catalog that contains photometry of sources over a wavelength range from 1.2 to 24~$\mu$m and is a combination of 2MASS/UKIDSS, Spitzer/IRAC, and MIPS data. The final catalogs for W43 and W51 are presented in Table~\ref{sourcetableW51} and \ref{sourcetableW43}, respectively.

\begin{deluxetable}{cccc}
\tabletypesize{\scriptsize}
\tablecolumns{3}
\tablecaption{Summary of Source Catalogs\label{datasummary}}
\tablewidth{0pt}
\tablehead{
\colhead{} & \colhead{} & \multicolumn{2}{c}{Number of Sources}\\
\cline{3-4}\\
\colhead{Telescope/Instrument} & \colhead{Band} & \colhead{W51} & \colhead{W43}}
\startdata
2MASS/UKIDSS & $J$ & 263,204 & 512,094 \\ 
2MASS/UKIDSS & $H$ & 272,935& 592,681 \\ 
2MASS/UKIDSS & $K_{s}$ & 270,380 & 606,648 \\ 
Spitzer/IRAC & 3.6 $\mu$m & 224,559 & 580,315  \\ 
Spitzer/IRAC & 4.5 $\mu$m & 189,793 & 540,343  \\ 
Spitzer/IRAC & 5.8 $\mu$m & 58,550 & 191,993 \\ 
Spitzer/IRAC & 8.0 $\mu$m & 42,372 & 126,838  \\ 
Spitzer/MIPS & 24.0 $\mu$m & 1505 & 3425 \\          
\enddata
\end{deluxetable}

\begin{deluxetable*}{lllcc}
\tablecolumns{4}
\tabletypesize{\scriptsize}
\tablewidth{0pt}
\tablecaption{Source-matching results\label{datasourcematch}}
\tablehead{
\colhead{} & \colhead{} & \colhead{} & \multicolumn{2}{c}{Number of matches\tablenotemark{a}}\\
\cline{4-5}\\
\colhead{Catalog A} & \colhead{Catalog B} & \colhead{$r_{match}$} & \colhead{W51} & \colhead{W43}}
\startdata
IRAC & 2MASS & 1$\arcsec$.0 & 214, 214 (75\%) & 370, 426 (58\%)  \\
IRAC+2MASS & UKIDSS & 1$\arcsec$.0 & 276, 324 (97\%) & 622, 867 (97\%)  \\
IRAC+(2MASS/UKIDSS) & MIPS & 1$\arcsec$.0 & 1505 (0.5\%) & 3425 (0.5\%)   \\
\enddata
\tablenotetext{a}{Match percentages are shown in parentheses.}
\end{deluxetable*}

\subsection{YSO Classification} \label{sec:yso}
\subsubsection{IRAC-color Criteria}\label{sec:iraccolor}
\begin{figure*}
\centering
\includegraphics[angle=270,width=16cm]{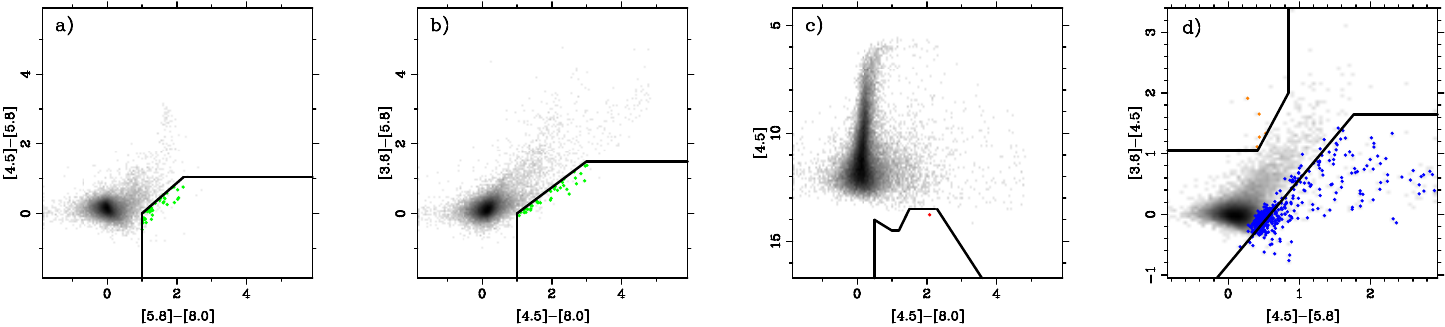}
\includegraphics[angle=270,width=16cm]{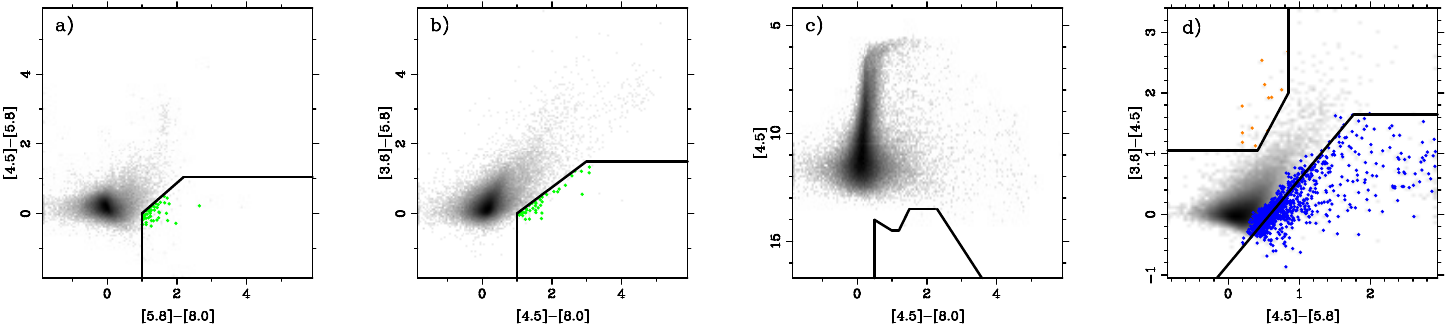}
\caption{Color$-$color diagrams used to identify contaminant objects among the sources with detection at all four IRAC bands following the criteria in \citep{gut09}. 
The background logarithmic gray-scale indicates the overall source density in each color-color space. The upper panels show the sources in W51, and the lower panels show the sources in W43. In panels (a), (b) PAH galaxies are marked with green circles. Panel c) shows the AGN galaxies (red circle) and panel d) shows knots of shocked emission (orange circles) and PAH-contaminated sources (blue circles).}\label{fig:irac4bandconts}
\end{figure*}

\begin{figure}
\centering
\includegraphics[angle=270,width=8cm]{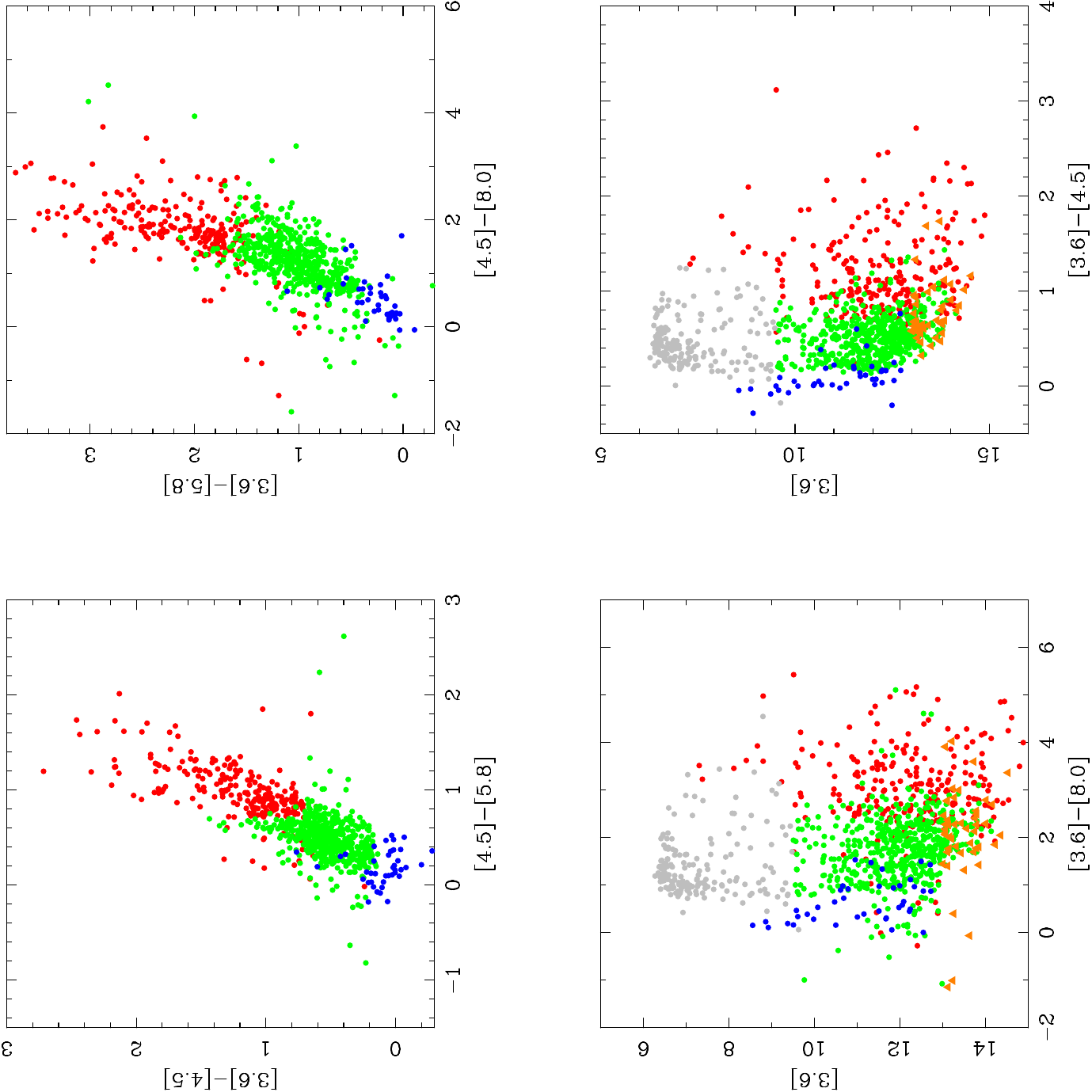}
\caption{Color$-$color and magnitude-color diagrams for the W51 region. Upper panel shows [4.5]-[5.8] vs. [3.6]-[4.5] and [4.5]-[8.0] vs. [3.6]-[5.8] IRAC color-color diagrams. Lower panel shows [3.6]-[8.0] vs. [3.6] and [3.6]-[4.5] vs. [3.6] IRAC magnitude-color diagrams. Red dots: class I; green: class II; blue: transition disk candidates; black: class III and photospheres; gray: AGB star candidates; orange triangles: background contaminants.}\label{fig:iracysosW51}
\end{figure}

\begin{figure}
\centering
\includegraphics[angle=270,width=8cm]{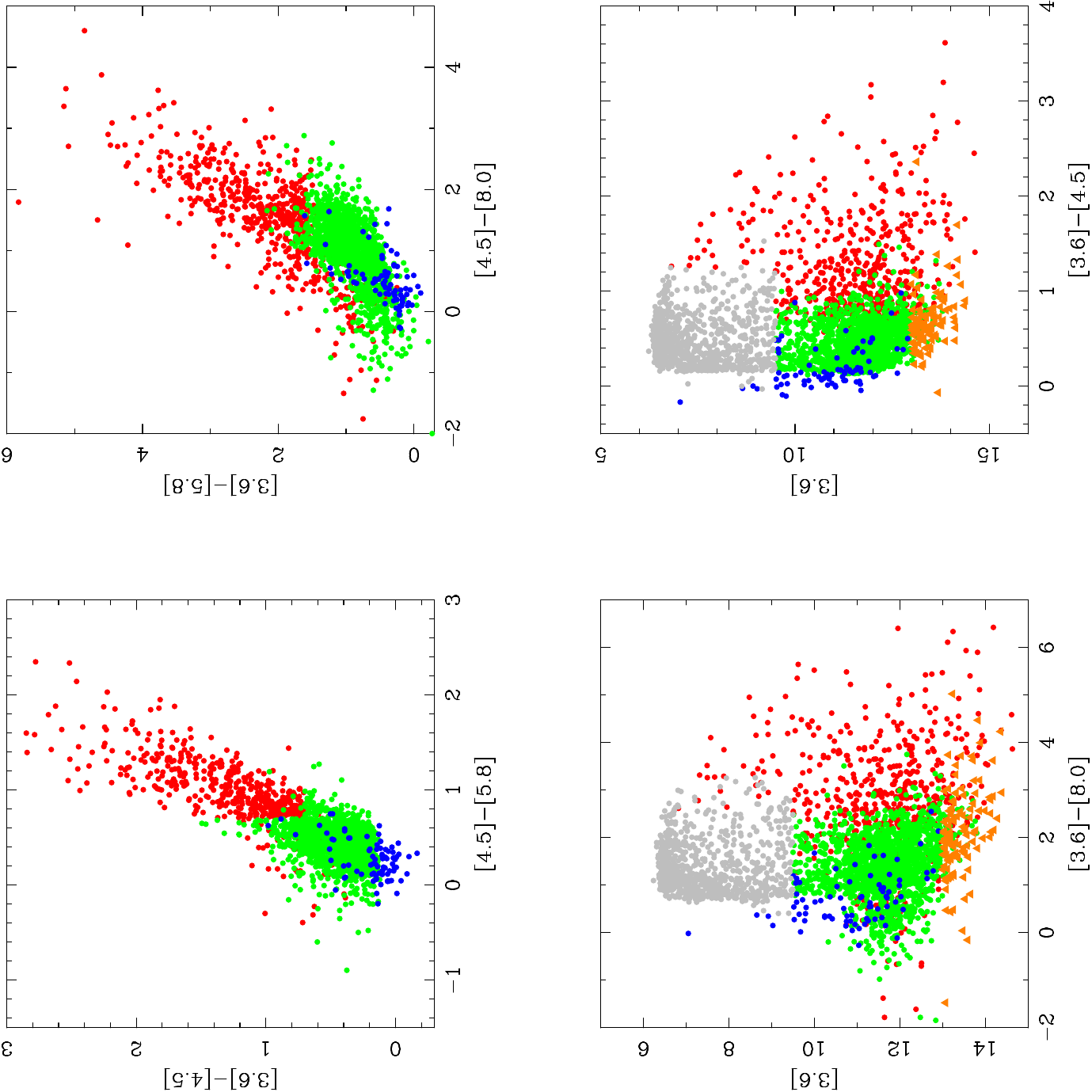}
\caption{Color-color and magnitude-color diagrams for W43 region. The upper panels show [4.5]-[5.8] vs. [3.6]-[4.5] and [4.5]-[8.0] vs. [3.6]-[5.8] IRAC color-color diagrams. The lower panels show [3.6]-[8.0] vs. [3.6] and [3.6]-[4.5] vs. [3.6] IRAC magnitude-color diagrams. Red dots: class I; green dots: class II; blue dots: transition disk candidates; black dots: class III and photospheres; gray dots: AGB star candidates; orange triangles: background contaminants.}\label{fig:iracysosW43}
\end{figure}

In this study we used the selection method based on color and magnitude criteria that we applied in \citet{sar15} for W49 and defined in \citet{gut08, gut09} (hereafter IRAC-color method). The \citet{gut09} IRAC-color method is updated to account for complications found in the more active and typically more distant regions. In Phase 1, by using all four IRAC bands we eliminated extragalactic sources such as star-forming galaxies and polycylic aromatic hydrocarbon (PAH) rich galaxies, since they can be classified from their very red  $5.8$ and $8.0$~$\mu$m colors \citep{ste05} due to strong PAH emission. In addition, we checked possible  broadline active galactic nuclei (AGNs) candidates that have consistent mid-IR colors with YSOs \citep{ste05} by using the [4.5]$-$[8.0] vs. [4.5] color-magnitude diagram as described in \citet{gut09}. In addition, we removed sources classified as knots of shock emission and PAH emission-dominated sources \citep{gut09}. As a result, we eliminated one AGN candidate, 74 PAH-rich galaxies, five knots of shocked gas emission, and 444 PAH-contaminated apertures in the W51 region and 86 PAH-rich galaxies, 13 knots of shocked gas emission, and 1317 PAH-contaminated apertures in the W43 region (Figure~\ref{fig:irac4bandconts}). No sources were classified as AGN candidates toward the W43 region. After eliminating contaminant sources, by using a combination of different color-magnitude conditions of the four IRAC bands described by \citet{gut09} we classified the YSOs into two categories: Class I sources (protostars with circumstellar disks and infalling envelopes), and Class II sources (pre-main-sequence stars with optically thick disks). Class I sources (shown with red color dots in Figure~\ref{fig:iracysosW51} and Figure~\ref{fig:iracysosW43}) are classified by their red colors using the following criteria \citep{gut09}:

\begin{displaymath}
[4.5]-[5.8] > 0.7 \text{ and}\  [3.6]-[4.5] > 0.7 
\end{displaymath}

Class II candidates were classified using the following constraints as described in \citet{gut09}, while $\sigma$$_{3}$ and $\sigma$$_{4}$ are the errors of the colors [4.5]$-$[8.0] and [3.6]$-$[5.8], respectively:

\begin{displaymath}
[4.5]-[8.0]-\sigma_{3} > 0.5
\end{displaymath}
\begin{displaymath}
[3.6]-[5.8]-\sigma_{4} > 0.35
\end{displaymath}
\begin{displaymath}
[3.6]-[5.8]+\sigma_{4} \le \frac{0.14}{0.04} \times (([4.5]-[8.0]-\sigma_{3})-0.5) + 0.5
\end{displaymath}
\begin{displaymath}
[3.6]-[4.5]-\sigma_{4} > 0.15
\end{displaymath}

With this method we can differentiate YSOs from field stars/photospheres \citep{ind05,fla07}. However, line-of-sight extinction and intrinsic variability can cause a reddened Class II source be misclassified as a Class I source. Therefore, the next step (Phase 2) in the \citet{gut09} method is to estimate the extinction ($A_{K}$) simply by using the measured $JHK$ magnitude and, when $J$ is not available, $3.6$ and $4.5$ ~$\mu$m data. The extinction was measured by using the baseline colors of Classical T Tauri Stars (CTTS) locus from \citet{mey97}, standard dwarf colors from \citet{bes88}, and baseline colors of YSO loci from \citet{gut05}. Following that the measured 1-8 ~$\mu$m magnitudes were dereddened according to the reddening law by \citet{fla07}. After separating the reddened Class II candidates from the Class I candidates, the third step (Phase 3) was to classify ``deeply embedded sources,'' which are Class I candidates with bright emission at 24~$\mu$m, and ``transition disk candidates,'' which are Class II sources with significant dust clearing within their disks which generates bright emission at 24~$\mu$m. Embedded protostar candidates are classified if they meet the following criteria:

\begin{displaymath}
[24] < 7 \text{ and}\ [X]-[24.0] > 4.5 
\end{displaymath}

where [X] is the longest wavelength detection with IRAC. Finally, transition disk candidates are classified by the following criteria:

\begin{displaymath}
[5.8] - [24] > 2.5 \text{ or}\ [4.5] - [24] > 2.5
\end{displaymath}

\begin{figure}
\centering
\includegraphics[angle=270,width=3.5cm]{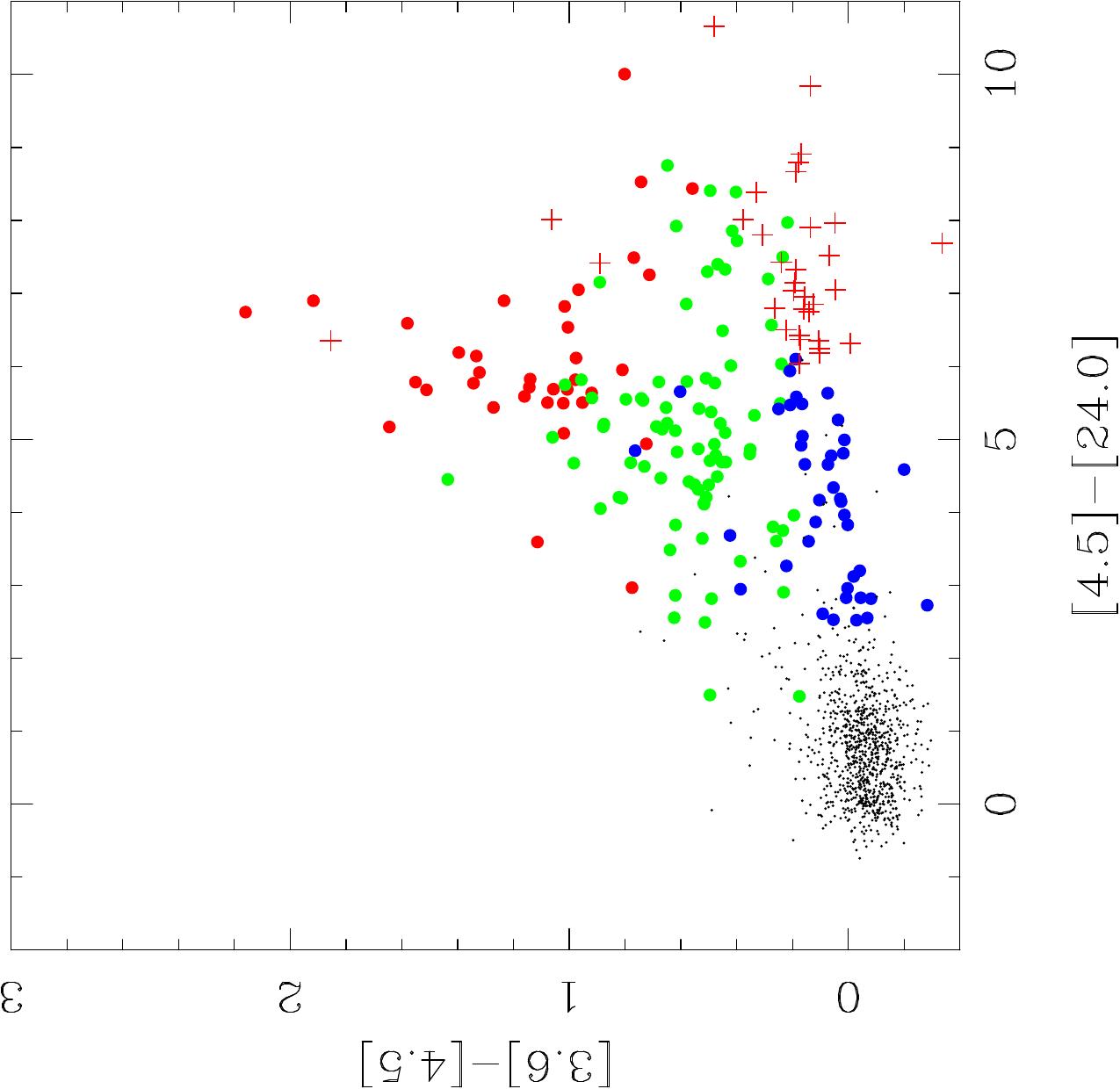}
\hspace{0.6cm}
\includegraphics[angle=270,width=3.5cm]{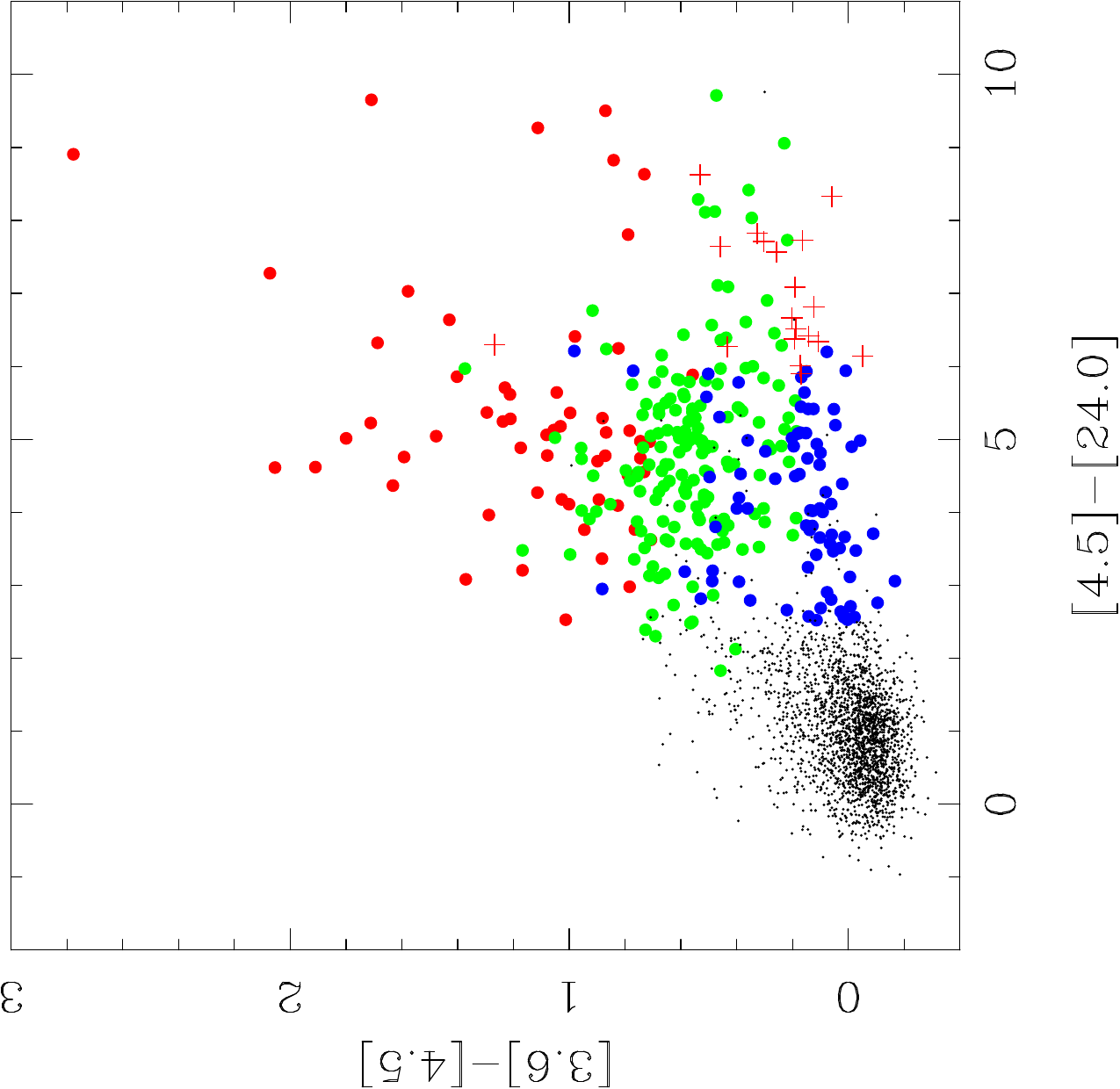}
\caption{[4.5]-[24.0] vs. [3.6]-[4.5] IRAC color-color diagram for W51 (left) and W43 (right). Red dots: class I; green dots: class II; blue dots: transition disk candidates; black dots: class III and photospheres; red asterisk: embedded protostars.}\label{fig:photospheres}
\end{figure}

Photospheres, transition disk candidatates, and embedded protostars can be seen in Figure~\ref{fig:photospheres}.

Following this method we also took into account the greater distance, and therefore the possible higher level of contamination, from objects such as AGB stars, background sources, and other extragalactic contaminants that have probably been classified as YSOs. Since AGB stars are bright and, in general, slightly bluer than YSOs \citep{gut08,koe14}, we reclassified the initially classified bright YSOs that also follow the following selection criteria as candidate AGB stars:

\begin{displaymath}
3.5 < [3.6] < 9.5 \text{ and}\ 0.4 < [3.6]-[8.0] < 2.6
\end{displaymath}
or 
\begin{displaymath}
3 < [3.6] < 9.5 \text{ and}\ 0.2 < [3.6]-[4.5] < 1.25 
\end{displaymath}

We classified 180 sources in the W51 region and 866 sources in the W43 region as candidate AGB stars, with their bright magnitudes shown in $[3.6]-[4.5]$ versus $[3.6]$ and $[3.6]-[8.0]$ versus $[3.6]$ color$-$magnitude diagrams (shown in the bottom panels of Figure~\ref{fig:iracysosW51} for W51 and Figure~\ref{fig:iracysosW43} for W43). The \citet{gut09} criteria require detections at $5.8$ or $8.0$~$\mu$m to identify extragalactic and background contaminants. However, we have only 37,932 sources in W51 out of 285,252 and 118,365 sources in W43 out of 641,732 that are detected in all four IRAC bands. It is therefore possible to have contaminant sources in our list of YSO candidates that have been classified with only IRAC $3.6$ and $4.5$~$\mu$m photometry. Thus, we applied a selection cut of [$3.6$] > 13 mag for the YSO candidates that are classified during the $JHK$[3.6][4.5] phase in order to separate faint YSO candidates from galaxies or background contaminants, and denoted the lower confidence in their identification by separating them from the rest of the YSO candidates. We identified 302 Class I YSOs and 1178 Class II/transition disk candidates in W51, and 917 Class I YSOs and 5187 Class II/transition disk candidates in W43. The difference in the YSO numbers is due to larger extent of W43. The color$-$color and color$-$magnitude diagrams of the identified YSOs, faint YSOs, and eliminated AGB star candidates are shown in Figure~\ref{fig:iracysosW51} and \ref{fig:iracysosW43}. The points are plotted without dereddening their photometry. In Table~\ref{sourcesum} we summarize the final source classification results. 

\citet{kan09} used \textit{Spitzer} and 2MASS data and classified 737 YSO candidates in W51 by using a combination of color criteria \citep{sim07} and SED slope criteria \citep{lad91,and93,gre94}. The class identification for the slope of log($\lambda$F$_{\lambda}$) vs. log($\lambda$) between $2$ and $\sim20$ $\mu$m is:

\begin{equation} \alpha\geq0.3 \textnormal{ Class I}\nonumber\end{equation}
\begin{equation}0.3 >\alpha\geq-0.3 \textnormal{ Flat Spectrum}\nonumber\end{equation}
\begin{equation}-0.3 >\alpha\geq-1.6 \textnormal{ Class II}\nonumber\end{equation}
\begin{equation}-1.6 >\alpha\geq-2.7 \textnormal{ Class III}\nonumber\end{equation}

They used mainly the SED slope method and not the preferred method in \citet{gut08}. However, the \citet{gut08} method has been slightly updated to better eliminate contaminant sources \citep{gut09}, and we have also applied an additional color selection method as explained above to eliminate more background contaminants and AGB candidates. We cross-matched our catalog with that of \citet{kan09} using a matching radius of 1\arcsec and found 91\% of the sources in common, the differences being due mainly to a slight difference of the cataloged region in both studies. Among the matched sources,  54\% (366/680) of the sources are identified as YSO candidates in both catalogs, where 123 of them are in the YSO clusters we describe in Section \ref{sec:clustering}. The rest of the sources are classified as contaminants or photospheres in our catalog. 17\% (117/680) of them are not classified in our catalog, because they lack photometry in at least four bands, so they appear as unclassified. However, \citet{kan09} classified the sources based on their SED slopes of the four IRAC bands or the SED slope of log($\lambda$F$_{\lambda}$) vs. log($\lambda$) between $2$ and $\sim24$ $\mu$m, without a constraint on the number of data points. 17\% (119/680) of \citet{kan09} sources are classified as contaminants in our study, such as AGBs, PAH galaxies, PAH-dominated sources, or faint or possibly background sources (classified as YSOs in our first classification phase but then eliminated based on the \citet{koe14} criteria). Finally, 12\% (80/680) of the \citet{kan09} YSOs were classified as photospheres in our catalog since they do not have the infrared color excess that is described above for Class I or Class II candidates.

\begin{deluxetable}{lcc}
\tablewidth{0pt}
\tablecolumns{3}
\tablecaption{Source Classification Summary\label{sourcesum}}
\tablehead{
\colhead{} & \multicolumn{2}{c}{Number of Objects}\\
\cline{2-3}\\
\colhead{Type} & \colhead{W51} & \colhead{W43}} 
\startdata
Class I & 302 & 917 \\
Class II & 1101 & 4993 \\
Transition Disks & 77 & 194 \\
Embedded protostars & 56 & 144  \\
Faint Class I\tablenotemark{*} & 100 & 303  \\
Faint Class II\tablenotemark{*} & 1618 & 4549 \\
Faint Transition Disk\tablenotemark{*} & 35 & 46  \\
Class III/Photospheres & 157, 165 & 429, 895 \\
AGB candidates & 180 & 866  \\
Other \tablenotemark{**} & 524 & 1416 \\
Unclassified & 124, 094 & 198, 409  \\
Total & 285, 252 & 641, 732 \\
\enddata
\tablenotetext{*}{These are the YSO candidates with lower confidence.}
\tablenotetext{**}{Includes PAH emission-dominated sources,
  H$_2$ shock emission dominated sources, broadline AGN
  candidates, and PAH galaxy candidates. Unclassified sources lack detection in four
  bands (either $HK_S$, IRAC 1 and 2, or IRAC 1, 2, 3, and 4) or a
  bright MIPS 24~$\micron$ detection.}
\end{deluxetable}

\subsubsection{SED Slope Distribution}

We calculated the SED slopes of the objects classified as YSOs using the IRAC-color criteria and determined their classifications using the SED slopes in order to compare the results from the two methods. 
We used all of the available photometry in the catalog except $J$ and $H$ and fit the slope of log($\lambda$F$_{\lambda}$) vs. log($\lambda$) (e.g., Figure~\ref{fig:W51_MYSO_SEDs}) between $2$ and $\sim20$ $\mu$m \citep{lad87,and93}, including the $24$ $\mu$m data, in order to make a comparison between IRAC-identified embedded sources and transition disk candidates. We calculated the flux densities by using zero-point flux densities of 666.7 Jy for F$_{Ks}$ \citep{coh03}, and 280.9, 179.7, 115.0, 64.13 (IRAC Instrument Handbook, Version 2.0.3), and 7.17 Jy (MIPS Instrument Handbook, Version 3.0) for F$_{3.6}$, F$_{4.5}$, F$_{5.8}$, F$_{8.0}$, and F$_{24}$, respectively. We used the isophotal wavelength values of 2.16 for [Ks], and 3.55, 4.44, 5.73, and 7.87 $\mu$m for [3.6], [4.5], [5.8], [8.0], respectively \citep{faz04}, and 23.68 $\mu$m for [24] (MIPSGAL Data Delivery Version 2.0). 

\begin{figure*}[ht!]
\centering
\includegraphics[width=7cm]{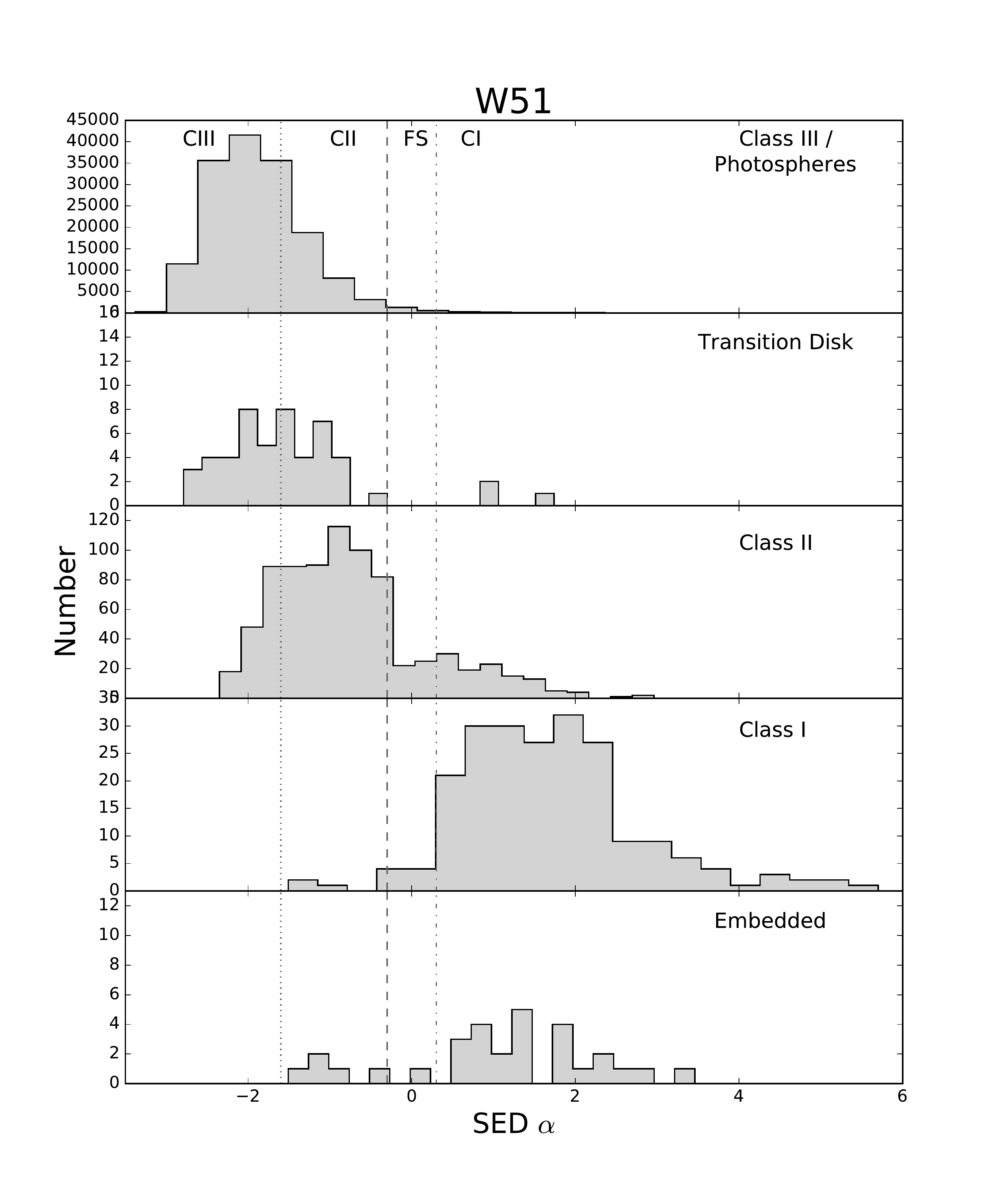}
\includegraphics[width=7cm]{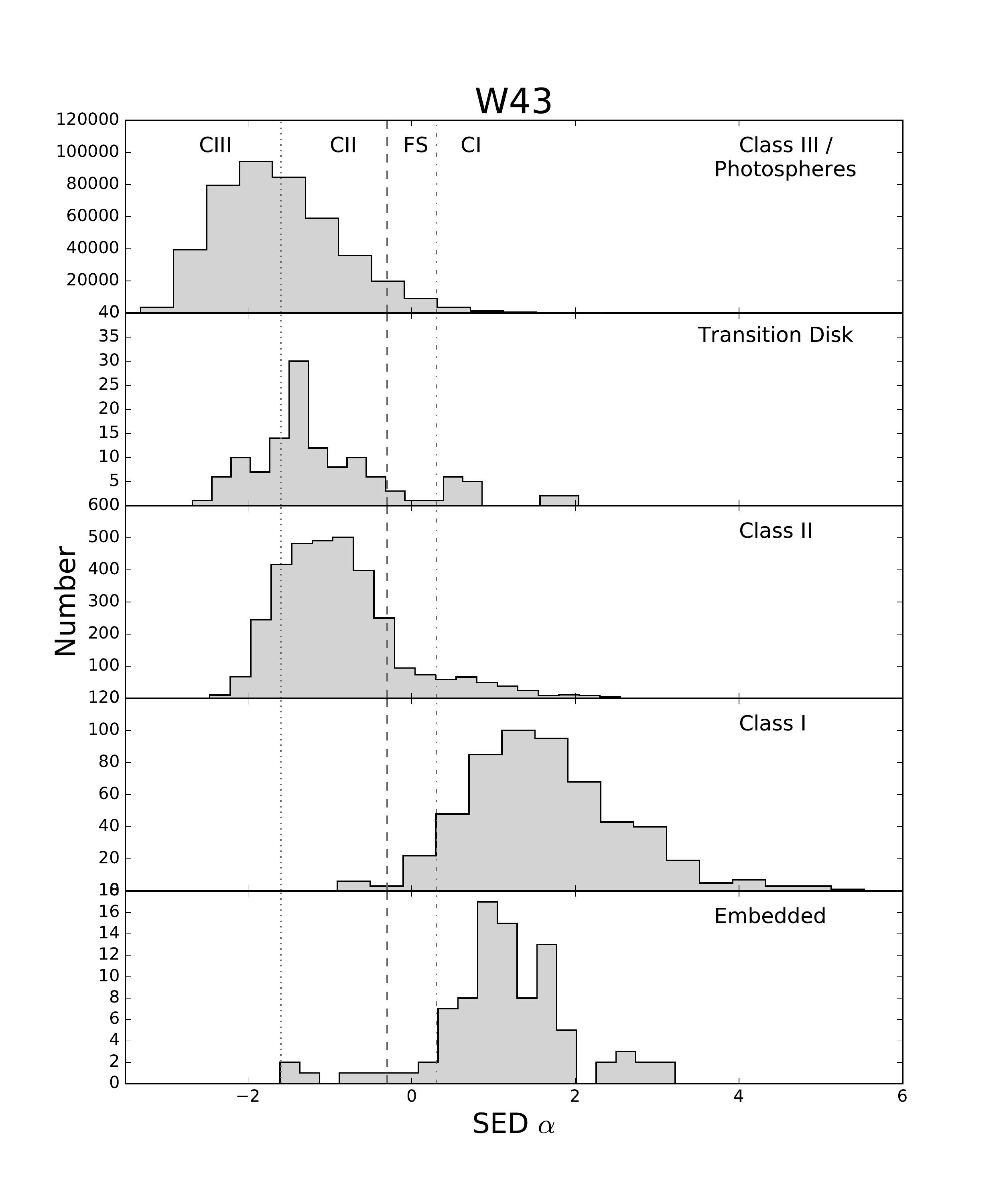}
\caption{The distribution of the SED spectral index $\alpha$ calculated from the original photometric data without correcting extinction in W51 (left panel) and W43 (right panel). The three vertical lines mark the division between the YSO regions based on their SED slopes (from left to right: Class III, Class II, Flat Spectrum, and Class I).}\label{fig:sedslopes}
\end{figure*}

The distribution of the SED spectral index for both regions (for sources with R-squared values of the linear fit of $>$0.8) can be seen in Figure~\ref{fig:sedslopes}. In both the IRAC-color and SED slope methods, the YSO classifications refer to the observational parameters rather than the evolutionary phases and the choices of subdivisions are considered arbitrary \citep[e.g.,][]{dun14}. When we compare the two, we find that the IRAC-color and SED slope methods give relatively consistent classifications for most objects. However, the comparison shows that using only the SED slope method does not help us to separate Class III or transition disk sources from Class II or embedded sources from Class I sources, while it does separate the Class III and Class II sources from the Class I or embedded sources. 

Looking at the data for the IRAC-color Class I objects in W43, 98\% of them have positive slopes as we expected. Similarly, we see that most of the embedded source candidates ($\sim$93\%) have positive slopes, indicating that they are young sources. For the IRAC-identified Class II candidates, $\sim$70\% of the SED slopes fall in the SED Class II range of -0.3 to -1.6. Most of them have negative slopes (89\% in W43), indicating that they are evolved sources, and the 11\% of them with positive slopes might be contaminant sources. Similarly, we see that 95\% of transition disk candidates (Class II sources with disks with inner gaps) are consistent with being evolved sources with negative slopes. Approximately 11\% of IRAC-identified transition disk candidates in W43 have positive slopes, which are most probably highly reddened transition disk candidates.

Overall, the histograms in Figure~\ref{fig:sedslopes} show that, except for a small percentage of sources, the SED slope distribution is consistent with the YSOs identified using the IRAC-color method. In order to see the effect of extinction correction on classification, we used the $A_{K}$ values that were estimated during the classification process in Section \ref{sec:iraccolor}, and dereddened the available photometry to calculate the SED slopes from $2$ to $24$ $\mu$m. While the extinction correction helps to reduce the small number of misclassifications of more evolved transition disk or Class II sources, it does not give a good result, especially for Class I or embedded sources; this is likely due to overcorrection. For IRAC-identified Class I candidates or embedded sources, extinction due to their envelopes and dusty disks plays an important role in the shape of their SEDs. Estimating an $A_{K}$ value and dereddening for these objects might be complicated, and it may cause misclassification due to their overcorrected colors and SED slopes. Indeed, the IRAC-color classification does not take into account the extinction for all sources except when double-checking Class II and Class I sources during Phase 2 if they are lacking detections at $5.8$ or $8.0$ $\mu$m); the comparison of IRAC-identified Class I sources according to the SED slopes without extinction correction gives more consistent results. The SED slope distribution in W51 yields similar results to W43, as can be seen in Figure~\ref{fig:sedslopes}. We provide the SED slope values that are calculated from the original photometric data with the errors and R-squared values in Table~\ref{sourcetableW51} and Table~\ref{sourcetableW43}. 

Both W51 and W43 are distant star-forming regions that contain significant amounts of high column density material, and young sources in these regions can be made extinct more easily by dust. In addition, there might be some misclassifications due to a mismatch of MIPS detections with their IRAC/near-IR counterparts. Due to the lower precision of MIPS astrometry, a transition disk candidate or a photosphere object can be classified as a deeply embedded source if it is mismatched with a MIPS source and lacks data in other bands. We have examined the catalog and images and removed obvious mismatched objects, but it is possible that chance alignments could lead to some that we cannot exclude with the available data.

\section{Clustering Analysis} \label{sec:clustering}

We used the Minimal Spanning Tree (MST) method \citep{caw04} to examine the substructure in the W51 and W43 regions, such as groups/clusters of YSOs, without using any kinematic information, as we did for W49 \citep{sar15}. The MST method identifies clusters as a collection of stars that are connected to each other by short branches (with a maximum of cutoff distance of $d_{c}$) with a minimum number of members ($N$) in a group. It has been used to identify groups/clusters of YSOs \citep{koe08,gut09,bee10,bil11,cha14,sar15} and dense cores \citep{kir16} for nearby star-forming regions, and also to identify OB associations in other galaxies \citep{pie01,bas07}.

There is no robust way to determine a branch length for the clustering determination. However, the most common way is to plot the distribution of branch lengths, fit straight lines through the long and short branch length domains, and choose the point of intersection (the Straight line Fit (SLF) method, see Figure 10 in \citet{sar15}). Choosing a branch length that falls between these two domains is also a good way to separate clusters from distributed sources \citep{gut08,gut09}. Recently, \citet{fis16} has identified the likely clusters in the Canis Major star-forming region by choosing the middle of the branch length range, which gave similar clusters.  W51 and W43 are distant regions, and due to resolution effects and lower sensitivity, multiple stars may blend into single objects and therefore fewer stars might be detected. This effect causes the branch length value in the SLF method to increase, causing increased sensitivity to the largest-scale structures within the cloud. Therefore, it is likely that there are more clustered sources that are unresolved at this distance \citep[e.g.,][]{lou14}.

To perform the clustering analysis we used a smaller field in both regions. This covers the main cloud structure but not the outer parts where there might be mainly line-of-sight contamination. The number of YSOs that we used for the clustering analysis is 857 and 5700 for W51, and W43, respectively. We used the MST method to perform a two-dimensional (2D) analysis by assuming that the projection effect would be small at these distances. However, in order to see the statistical significance of the identified MST groups in the 2D analysis, we performed Monte Carlo simulations of uniformly random distributed objects (see Section \ref{sec:random}). \citet{sch06} also studied the effect of the 2D projections of the three-dimensional (3D) model clusters and concluded that although the clustering parameters can be different (and still giving similar mean values), the general behavior of the evolution of clusters is independent of the projection. 

\subsection{Clustering Results in W51}

We first determined a branch length by using the SLF method, plotting the number of clustered objects versus branch lengths using a standard value of $N$~$\geq$~10 for the minimum number of cluster members. Using a branch length of 82$\arcsec$.5 (which corresponds to 2.17 pc at the distance of W51), we found nine groups that contain 62\% (536/857) of the total number of YSOs in the region. In order to investigate how clustering properties change with the requirement of minimum members in a group, we repeated the analysis with a mix of 5 to 15 members for a branch length of 82$\arcsec$.5. While we identified 26 MST groups with $N$~$\geq$~5 and nine MST groups with $N$~$\geq$~10, we identified only five MST groups with $N$~$\geq$~15. When we decreased the number of minimum members from 15 to 5, we identified new small groups but the five largest MST groups remained unchanged.  

\begin{figure*}
\includegraphics[width=9cm]{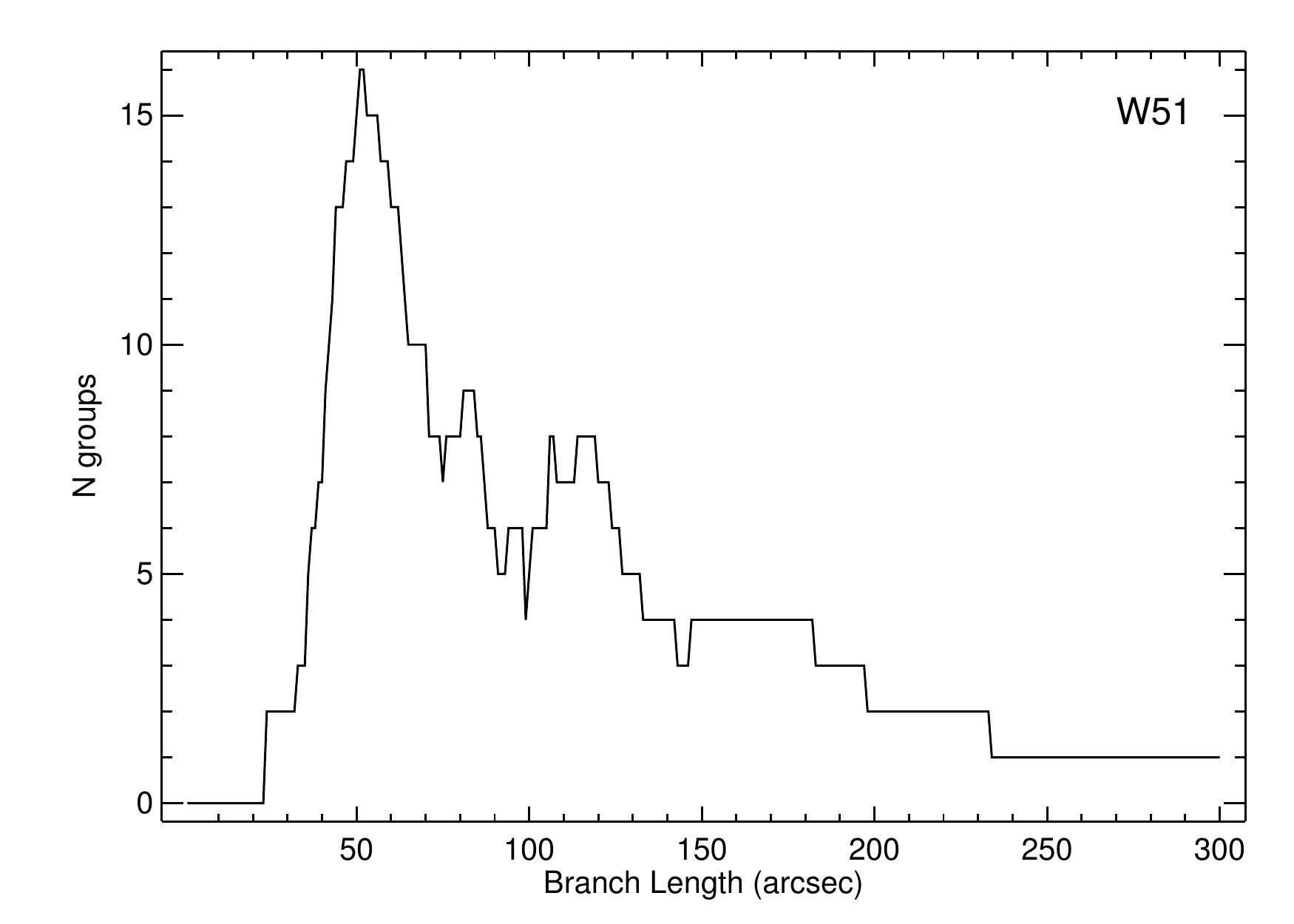}
\includegraphics[width=9cm]{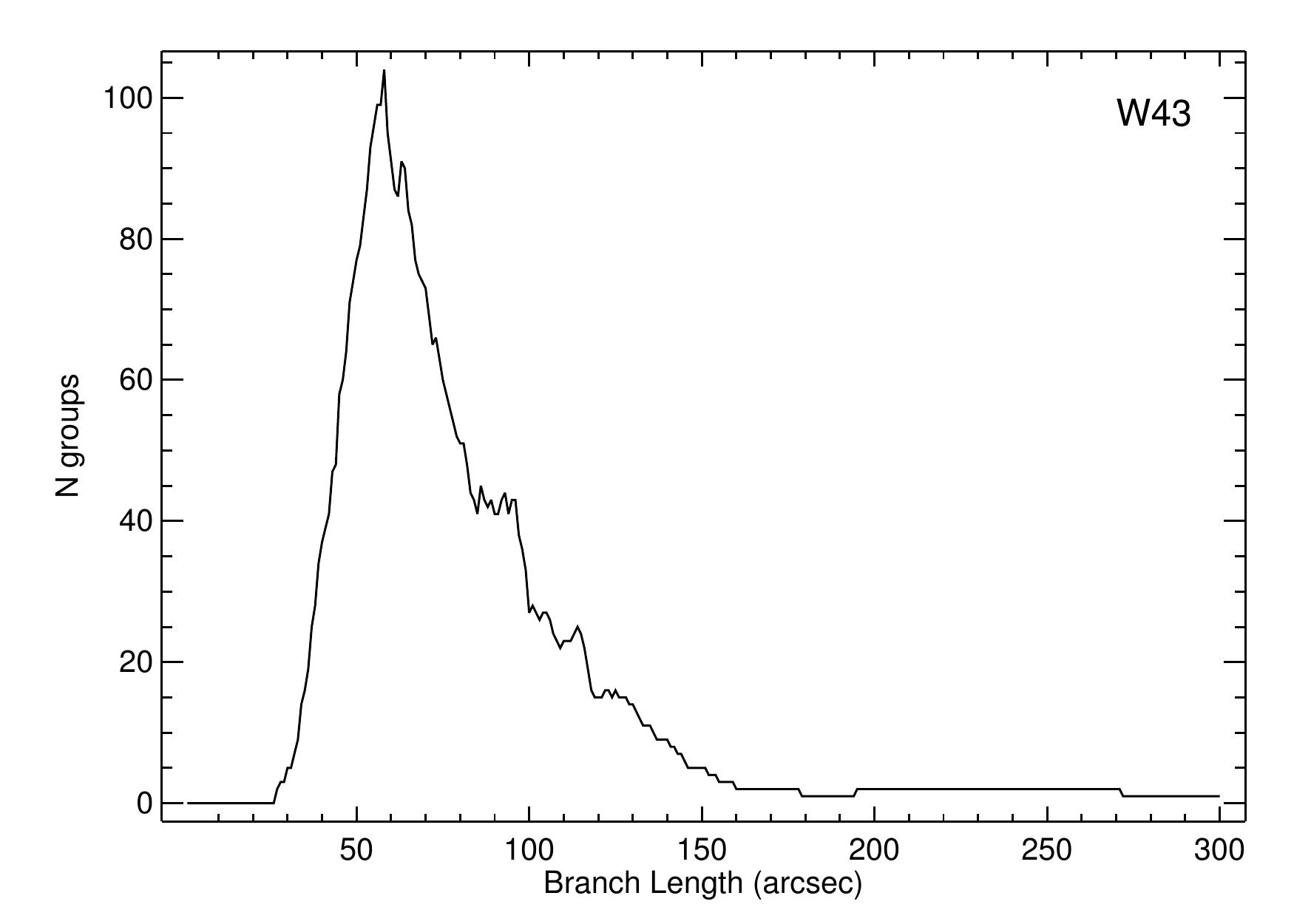}
\caption{Number of groups containing 10 or more stars identified by the MST algorithm. The left panel is plotted for the W51 region, and the right panel is plotted for the W43 region. \label{fig:SLF}}
\end{figure*}

\begin{figure*}
\centering
\includegraphics[width=18cm]{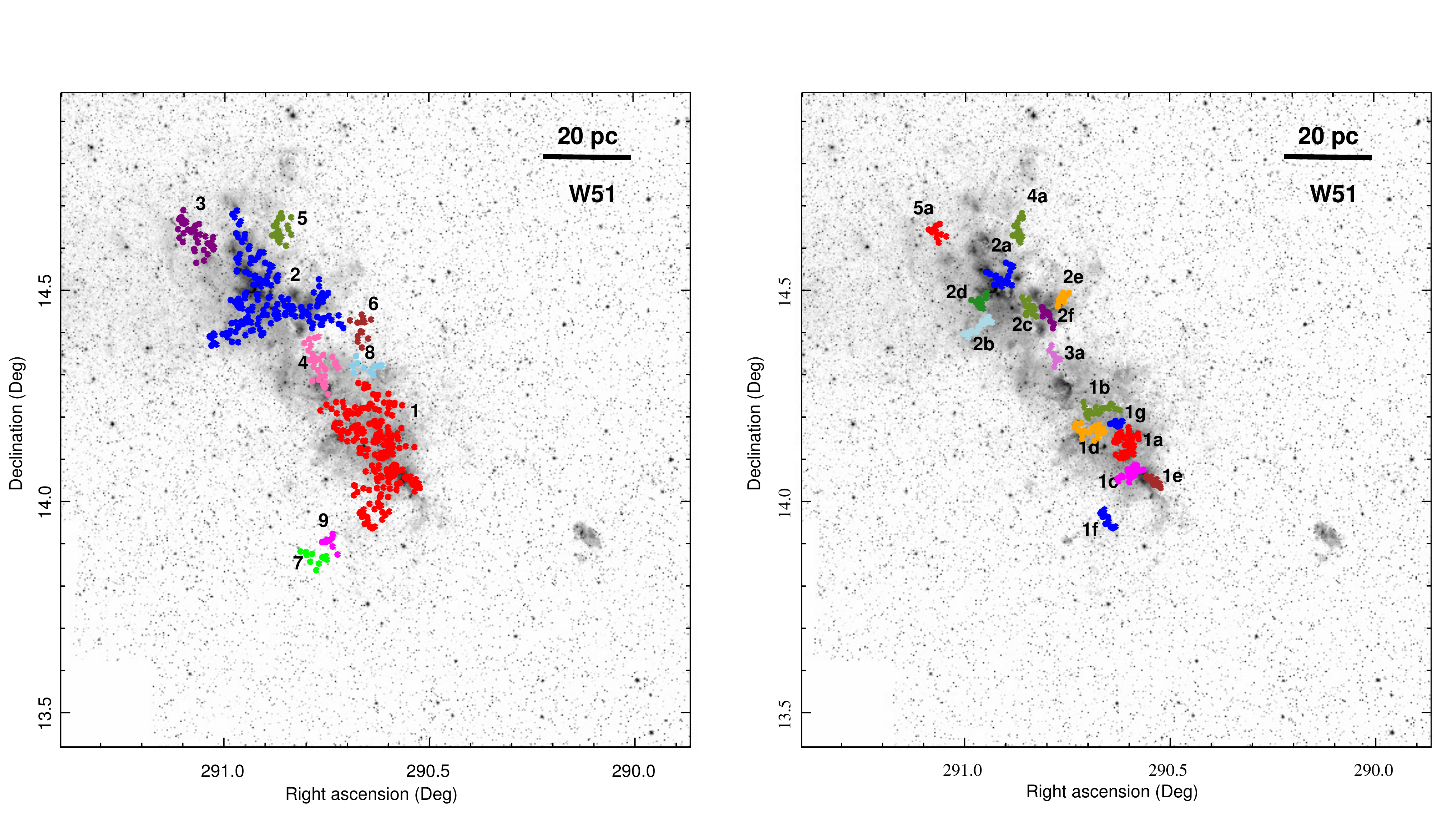}
\caption{Left: the nine MST groups identified using $d_c$ = 82$\arcsec$.5 are shown with group numbers overlaid on the IRAC 4.5~$\mu$m image. Right: the 16 MST subgroups identified by $d_c$ = 51$\arcsec$ are shown with a cluster number and a letter (see Table~\ref{W51GMCclusters}).}\label{fig:W51mstresults2}
\end{figure*}

To investigate how the numbers of identified groups change with different branch lengths, we plotted the number of groups containing 10 or more stars against cutoff distances from 1$\arcsec$ to 300$\arcsec$, with steps of 1$\arcsec$, in the left panel of Figure~\ref{fig:SLF}. A maximum of 16 MST groups can be identified using a branch length of 51$\arcsec$ (1.34 pc), while 33\% (286/857) of the YSO candidates are associated with a cluster. The MST groups identified by both branch length values can be seen on the upper panel of Figure~\ref{fig:W51mstresults2} and their properties are summarized in Table~\ref{W51GMCclusters}. The results show that using a shorter branch length allows us to identify subgroups and see the hierarchical structure. We also investigated the effects of changing the requirement of minimum members for 51$\arcsec$ and we identify 33 MST groups with $N$~$\geq$~5, 16 groups with $N$~$\geq$~10, and nine groups with $N$~$\geq$~15. As expected with the increasing number of minimum members, we continue to identify the same largest groups, but lose the smaller ones. Our clustering analysis shows that the MST groups in W51 have 10$-$217 sources with diameters of 2.5$-$30~pc, and the subgroups within them have 10$-$44 sources with diameters of 1$-$7~pc (Table~\ref{clustersumW51}). 

\begin{deluxetable*}{lccccccccccc}
\tabletypesize{\scriptsize}
\tablecaption{Properties of the Clusters in W51 \label{W51GMCclusters}}
\tablewidth{0pt}
\tablehead{\colhead{No.\tablenotemark{a}} & \colhead{R.A. (J2000.0)\tablenotemark{b}} & \colhead{Decl. (J2000.0)\tablenotemark{b}} & \colhead{N$_{IR}$\tablenotemark{c}} & \colhead{I\tablenotemark{d}} & \colhead{II\tablenotemark{d}} & \colhead{II/I\tablenotemark{e}} & \colhead{N$_{emb}$\tablenotemark{f}} & \colhead{N$_{td}$\tablenotemark{g}} & \colhead{N$_{MYSO}$} & \colhead{Diameter} & \colhead{Diameter}\\
\colhead{} & \colhead{($\degr$)} & \colhead{($\degr$)} & \colhead{}  & \colhead{}  & \colhead{}  & \colhead{}  & \colhead{} & \colhead{} & \colhead{} & \colhead{($\arcsec$)} & \colhead{(pc)}}
\startdata
1 & 290.628161  &  14.121185    &   217    &    53    &  158   &  2.98(0.48) & 4 & 2 & 5 & 1178.69  & 30.94 \\
1a & 290.607449 &  14.137096    &    44    &     9    &   35  &   3.89(1.50) & 0 & 0  & 1 & 162.35   &  4.26 \\
1b & 290.671961 &  14.218522    &    21    &     8    &   12  &   1.50(0.71) & 0    &     1  & 1 & 238.02   &  6.25 \\
1c & 290.595251 &  14.069474    &    20    &     4    &   16  &   4.00(2.41) & 0    &     0  & 0 & 159.93   &  4.20 \\
1d & 290.693620 &  14.168545    &    19    &     3    &   15  &   5.00(3.49) & 1    &     0  & 1 & 55.81     &  1.46 \\
1e & 290.539286 &  14.046652    &    16    &     9    &    7  &   0.78(0.41) & 0    &     0  & 1 & 102.86   &  2.70 \\
1f & 290.655596 &  13.961203    &    14    &     6    &    8  &   1.33(0.76) & 0    &     0  & 0 & 95.26     &  2.50 \\
1g & 290.629889 &  14.185851    &    14    &     2    &   12  &   6.00(5.29)  & 0    &     0  & 0 & 39.95     &  1.05 \\
2 & 290.895659  &  14.483122    &   175    &   55    &  107   &  1.94(0.32)  &  9    &     4 &  8 & 1254.22 &  32.92 \\
2a & 290.911695 &  14.532360    &    21    &     7    &   13  &   1.86(0.91)   &     1    &     0  & 2 & 255.73  &   6.71\\
2b & 290.968399 &  14.413716    &    18    &     5    &   10  &   2.00(1.16)   &   0    &     3  & 0 & 87.22    &   2.29\\
2c & 290.844332 &  14.459009    &    15    &     3    &   10  &   3.33(2.41)   &    1    &     1  & 0 & 88.78    &   2.33\\
2d & 290.962397 &  14.474087    &    12    &     3    &    9  &   3.00(2.20)   &      0    &     0  & 0 & 59.63    &   1.56\\
2e & 290.798509 &  14.440177    &    11    &     7    &    4  &   0.57(0.39)   &     0    &     0  & 0 & 85.10    &   2.23\\
2f & 290.762695 &  14.480864    &    11    &     5    &    5  &   1.00(0.68)   &        1    &     0  & 4 & 55.42    &   1.45\\
3 & 291.075362  &  14.632441    &    37    &    8    &   24   &  3.00(1.27)   &    3    &     2 &  0 & 175.98  &   4.62 \\
3a & 291.071001 &  14.638581    &    11    &     2    &    9  &   4.50(4.05)   &   0    &     0  &  0 & 137.68  &  3.61\\
4 & 290.767228  &  14.327158    &    35    &   10    &   22   &  2.20(0.86)   &      3    &     0 &  1 & 111.64  &   2.93 \\
4a & 290.779470 &  14.342320    &    12    &     7    &    5  &   0.71(0.45)   &  0    &     0  &  0 & 214.01 &  5.62\\
5 & 290.863964  &  14.644746    &    26    &    9    &   12   &  1.33(0.61)   &  5    &     0 &   1 & 103.76 &  2.72 \\
5a & 290.868706 &  14.645758    &    20    &     7    &    9  &   1.29(0.68)   & 4    &     0  &  0 & 277.27 &  7.28\\
6 & 290.665391  &  14.411405    &    13    &    1    &   11   & 11.00(14.95)  &     0    &     1 &  0 & 114.24 &   3.00 \\
7 & 290.776093  &  13.867202    &    12    &    0    &    8   &  \nodata      &     3    &     1 &   0 & 96.51   &  2.53 \\
8 & 290.656547  &  14.319608    &    11    &    3    &    7   &  2.33(1.77)   &     1    &     0 &   0 & 216.18  & 5.67 \\
9 & 290.743948  &  13.905521    &    10    &    3    &    6   &  2.00(1.55)   &      0    &     1 &  1 & 258.82  &  6.74 \\
\enddata
\tablenotetext{a}{Subgroups identified with $d_{c}$ = 51$\arcsec$ are named under the main group with main number and a letter.}
\tablenotetext{b}{Central coordinates of the YSO groups or subgroups.} 
\tablenotetext{c}{Number of stars with infrared excess. Includes Class I, Class II, deeply embedded protostars, and transition disk candidates.} 
\tablenotetext{d}{I; Class I, II; Class II candidates}
\tablenotetext{e}{Number in parentheses indicates uncertainty in ratio.} 
\tablenotetext{f}{Number of deeply embedded protostar candidates.} 
\tablenotetext{g}{Number of transition disk candidates.}
\end{deluxetable*}

\subsection{Clustering Results in W43}

We identified the MST groups and subgroups within the W43 region by using the same method. We first determined a branch length with the SLF method by using standard value of $N$~$\geq$~10 for minimum number of cluster members. 51 MST groups were identified with a branch length of 81$\arcsec$ (which corresponds to 2.20 pc at the distance of W43) and a total of 73\% (4176/5700) of YSOs are associated with a cluster. We identified 134 MST subgroups with $N$~$\geq$~5, 51 groups with $N$~$\geq$~10, and 27 groups with $N$~$\geq$~15. These 27 largest MST groups remained unchanged, while we identify new small groups with the decreasing number of minimum members.   

As we did for W51, we investigated the change in the number of clusters with different branch lengths (see the right panel of Figure~\ref{fig:SLF}). A maximum of 95 MST groups can be identified with a branch length of 59$\arcsec$ (1.61 pc), while 43\% (2448/5700) of the YSO candidates are associated with a cluster. The MST groups identified by both branch length values can be seen in Figure~\ref{fig:MSTclustersW43} and the properties of the five largest groups are summarized in Table~\ref{W43GMCclusters}. We identified 242 MST subgroups with $N$~$\geq$~5, 95 MST groups with $N$~$\geq$~10, and 53 MST groups with $N$~$\geq$~15 for 59$\arcsec$; the largest 53 MST groups remained unchanged with any chosen minimum number of group members. Our clustering analysis shows that identified MST groups in W43 have 10$-$3047 sources with diameters of 1.4$-$142~pc and subgroups within them have 10$-$217 sources with diameters of 0.8$-$14~pc (Table~\ref{clustersumW43}).  

\subsection{Random Clustering Analysis}\label{sec:random}

To determine the statistical significance of the identified MST groups, we performed Monte Carlo simulations of uniformly random distributed objects in order to examine their clustering properties and compare them with the observed ones. We created 1000 realizations of 857 objects spread randomly in a $0.5\times1.3$ degree field to match the number of YSOs and the size of the W51 region. We used both the branch length determined from the MST method and the branch length giving the maximum number of groups (subgroups) in order to compare both results with the observed groups. The largest random group among all of the simulations has 58 members, with an average maximum group size of $24\pm6$ members. In comparison, the observed largest group in W51 is significantly larger, with 215 members. The second largest random group in each of the simulations has on average $19\pm4$ members, with a  maximum of 36,  while the observed second largest group is still larger, with 172 members. Similarly, the average sizes of third, fourth, and fifth largest groups in each simulation are smaller than the observed ones. However, the smaller observed groups (sixth to ninth largest) have similar sizes to the average size of random groups, which indicates that they might be random associations. For the subgroups identified with the branch length of 51$\arcsec$, we see that all of the observed subgroups are larger than the mean size of random subgroups ($8\pm4$), which indicates that they are different than random distributions. Similarly, we generated 1000 simulations of 5700 randomly distributed sources in a $2.8\times2.1$ degree field, which is the size of the W43 region. Almost all of the observed groups and subgroups are larger than the average random group size, ($8\pm4$). 

\begin{figure*}
\centering
\includegraphics[width=8cm]{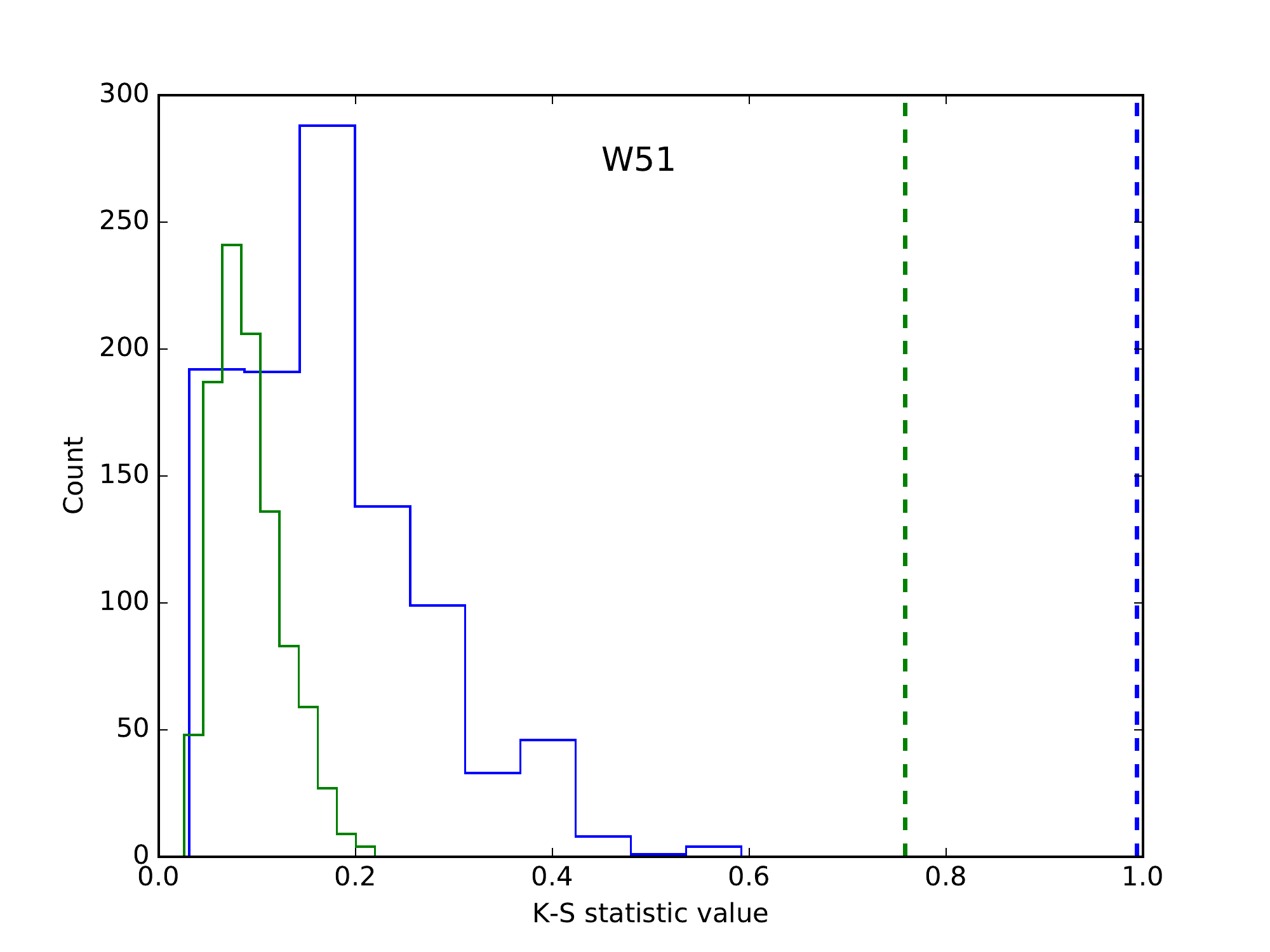}
\includegraphics[width=8cm]{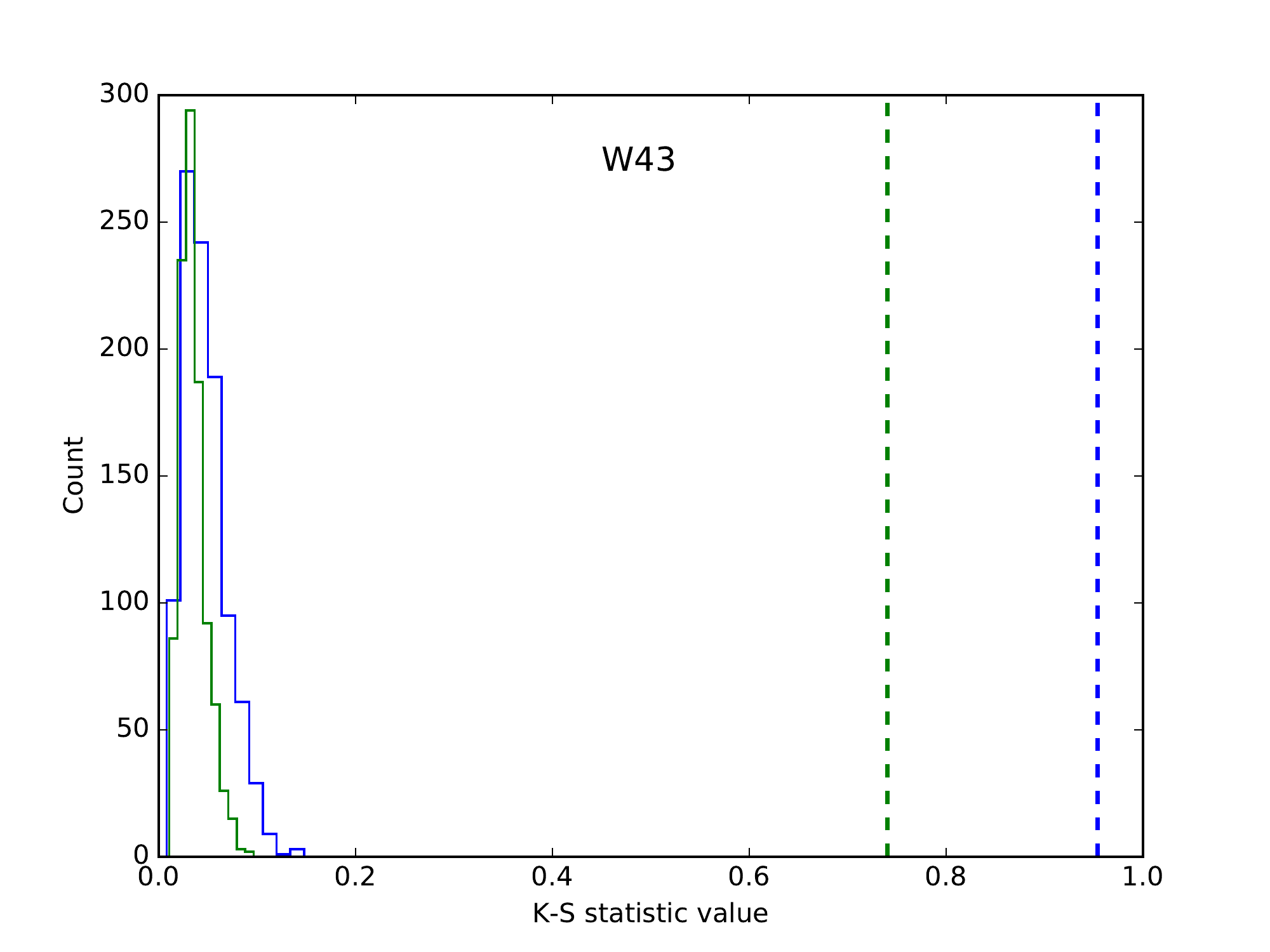}
\caption{K$-$S statistic distribution of random MST groups, between each of the 1000 random realizations and the mean random cluster distribution. The green histogram shows the groups identified with SLF branch length $d_c$ = 82$\arcsec$.5 in the W51 region (left), and $d_c$ = 81$\arcsec$ in the W43 region (right). The blue histogram shows the groups identified with branch length $d_c$ = 51$ \arcsec$ in the W51 region (left), and $d_c$ = 59$ \arcsec$ in the W43 region (right). Dashed lines in both figures represent the K$-$S statistic value between the mean CDF of all random clusters and the CDF of observed MST groups, with the same color code of histograms. For both regions, the inferred probability that the observed MST groups are drawn from a random distribution is less than $10^{-3}$.}\label{fig:KSresult}
\end{figure*}

\begin{figure*}
\includegraphics[width=9cm]{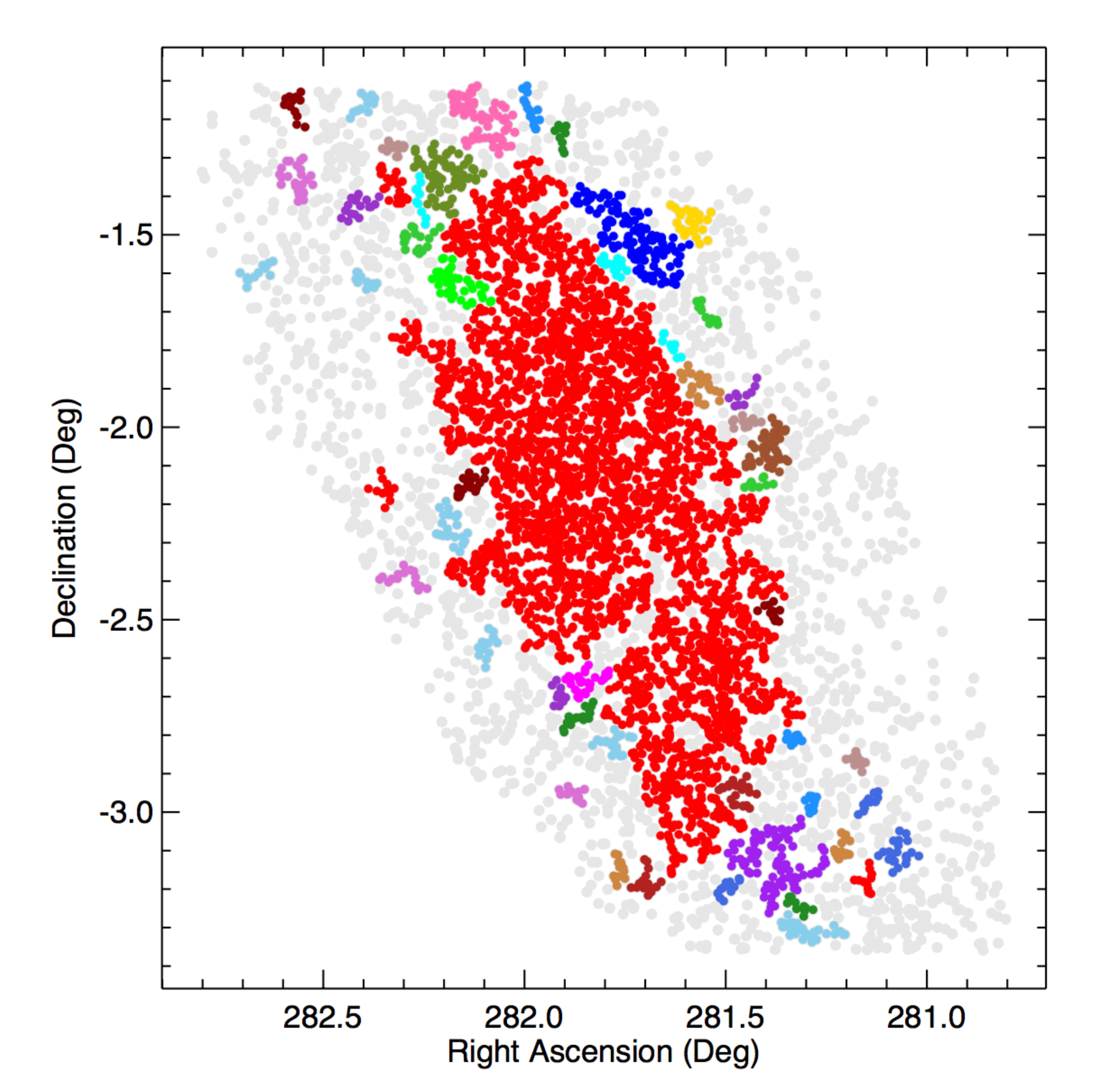}
\includegraphics[width=9cm]{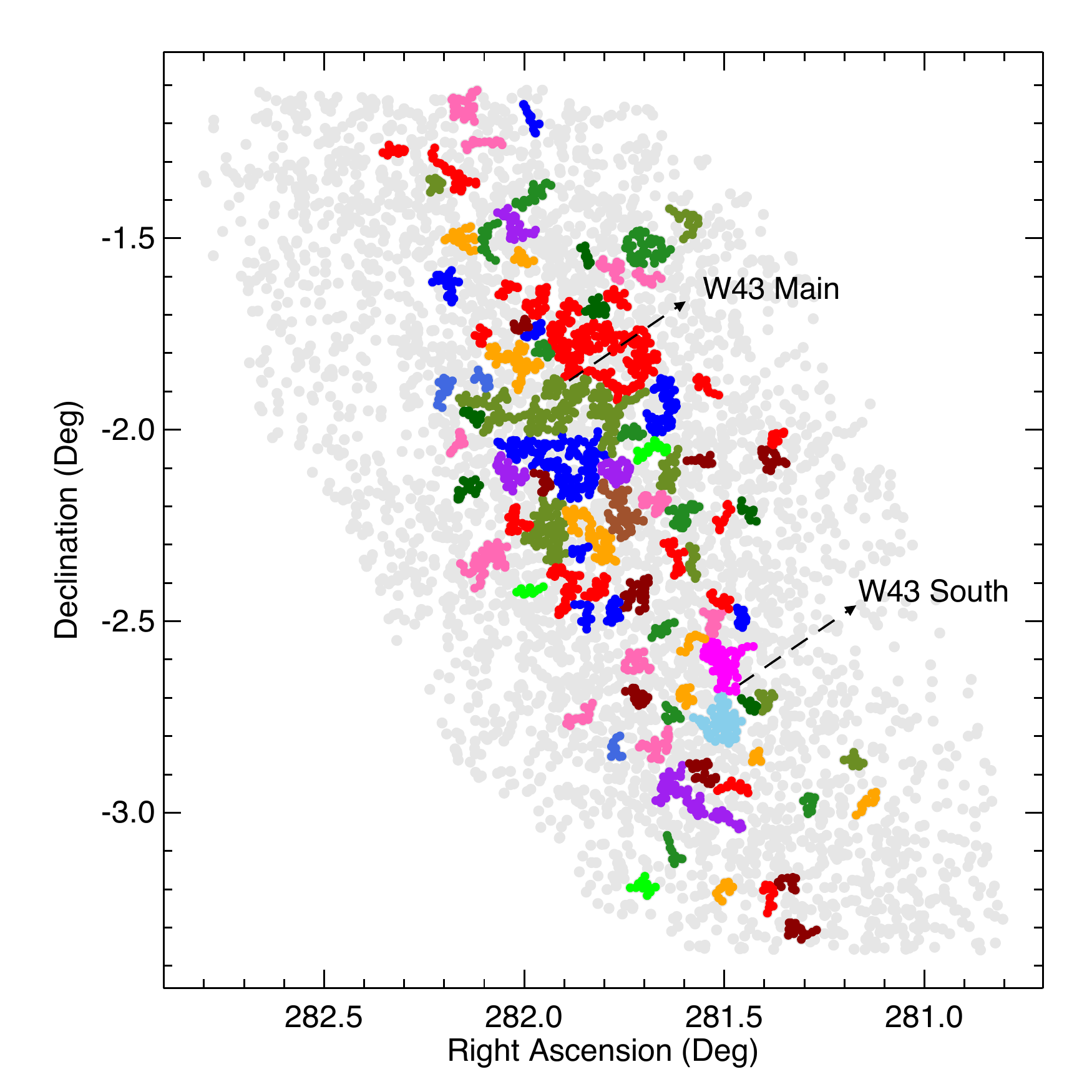}
\caption{Left: the 51 MST groups determined using the SLF method with a branch length of  81$\arcsec$ (2.20 pc for objects at the distance of W43). Right: the 95 MST subgroups determined with a branch length of 59$\arcsec$ (1.61 pc). The YSOs are plotted in colors according to the identified clusters. \label{fig:MSTclustersW43}}
\end{figure*}

As a final step we used a two-sided Kolmogorov-Smirnov (K-S) test \citep{con99} to determine if the two datasets (observed and random clusters) differ significantly. We calculated the two-sided K-S statistic between each of the 1000 realizations of randomly distributed sources and the mean cumulative distribution function (CDF) of random clusters from all 1000 realizations together. The distribution of the K-S statistic values, for the branch lengths of $d_c$ = 82.5$\arcsec$ and $d_c$ = 51$\arcsec$, are shown in Figure~\ref{fig:KSresult} as green and blue histograms, respectively. In the same figure, the K-S statistic value, between the CDF of observed MST groups and the mean CDF of random clusters, is shown with a vertical dashed line. We find that for both branch lengths, the K-S statistic value between the observed group size CDF and the mean random one is well outside of the distribution of K-S statistic values between the mean random CDF and  those of each of the 1000 realizations, Thus, we infer that the probability of the distribution of a real MST group being drawn from a random distribution is less than $10^{-3}$.

\begin{deluxetable*}{lccccccccccc}
\tabletypesize{\scriptsize}
\tablecaption{Properties of the Clusters in W43 \label{W43GMCclusters}}
\tablewidth{0pt}
\tablehead{\colhead{No.\tablenotemark{a}} & \colhead{R.A. (J2000.0)\tablenotemark{b}} & \colhead{Decl. (J2000.0)\tablenotemark{b}} & \colhead{N$_{IR}$\tablenotemark{c}} & \colhead{I\tablenotemark{d}} & \colhead{II\tablenotemark{d}} & \colhead{II/I\tablenotemark{e}} & \colhead{N$_{emb}$\tablenotemark{f}} & \colhead{N$_{td}$\tablenotemark{g}} & \colhead{N$_{MYSO}$} & \colhead{Diameter} & \colhead{Diameter}\\
\colhead{} & \colhead{($\degr$)} & \colhead{($\degr$)} & \colhead{}  & \colhead{}  & \colhead{}  & \colhead{}  & \colhead{} & \colhead{} & \colhead{} & \colhead{($\arcsec$)} & \colhead{(pc)}}
\startdata
1       &  281.799575         & -2.183919   		&  		3047    	&  	561   	&  	2358   	&  	4.20(0.20)     		&  	73    &   	55    & 11  &  412.90  & 142.07 \\
1*      &  281.806380 	& -1.792551   		&   		217 		&       59		&      143		&      2.42(0.38)		         &     10    &        5     & 1   &   443.85  & 11.65\\
2*  	&  281.912313 		& -2.081571   		&   		140    	&       36   		&      101   	&      2.81(0.55)     		&       2    &         1    & 3    &  554.08  & 14.54\\
...     &      \nodata 		&      \nodata		&      		\nodata		&      \nodata	&      \nodata	&      \nodata			&      \nodata &      \nodata &      \nodata &      \nodata &      \nodata	\\
2      &  281.726876		& -1.506898	        &  		 127	        &  	 20 		&      100 	        &       5.00(1.24)	        &	 3    &    4  	& \nodata &  209.98  & 31.76\\
12* 	&  281.695879 		& -1.525841   		&    		49    		&   12   		&      35   		&  2.92(1.00)    			 &   2    	&    0  	& \nodata &  247.25  &  6.49\\
61*	 &  281.769341 	& -1.654018   		&    		13    		&    5   		&       8   		&  1.60(0.97)		        &      0    	&    0  	& \nodata &  160.54  &  4.21\\
... & \nodata 		&      \nodata		&      \nodata		&      \nodata	&      \nodata	&      \nodata			&      \nodata &      \nodata &      \nodata &      \nodata &      \nodata	\\
3   &  281.379116 & -3.130973   &    86    &    8   &    72   &  9.00(3.49)     &   2    &    4  & \nodata &  516.63  & 13.56\\
.... & \nodata 		&      \nodata		&      \nodata		&      \nodata	&      \nodata	&      \nodata			&      \nodata &      \nodata &      \nodata &      \nodata &      \nodata	\\
\enddata
\tablecomments{Table 7 is published in the electronic edition of the {\it Astrophysical Journal}. A portion is shown here for guidance regarding its form and content.} 
\tablenotetext{a}{Subgroups identified with $d_{c}$ = 59$\arcsec$ are named under the main group with main number and an asterisk.}
\tablenotetext{b}{Central coordinates of the YSO groups or subgroups.} 
\tablenotetext{c}{Number of stars with infrared excess. Includes Class I, Class II, deeply embedded protostars, and transition disk candidates.} 
\tablenotetext{d}{I; Class I, II; Class II candidates}
\tablenotetext{e}{Number in parentheses indicates uncertainty in ratio.} 
\tablenotetext{f}{Number of deeply embedded protostars.} 
\tablenotetext{g}{Number of transition disk candidates.}
\end{deluxetable*}

\begin{deluxetable}{lll}
\tabletypesize{\scriptsize}
\tablecaption{Clusters/Subgroups in W51 \label{clustersumW51}}
\tablewidth{0pt}
\tablehead{\colhead{Parameter} & \colhead{SLF} & \colhead{$d_c$ = 51$\arcsec$}\\
\colhead{} & \colhead{Method} & \colhead{}}\\
\startdata
Number of clusters & 9 & 16\\
Cutoff distance & 82$\arcsec$.5(2.17~pc) & 51$\arcsec$(1.34~pc)\\
Percent in clusters & 60 & 32\\
Group diameter & 96$\arcsec$.5-1178$\arcsec$ (2.5-30~pc) & 40$\arcsec$-256$\arcsec$ (1-7~pc)\\
Class II/I Ratio & 2.50(0.25)\tablenotemark{*} & 2.06(0.27)\tablenotemark{*}\\
\enddata
\tablenotetext{*}{Number in parentheses indicates uncertainty in ratio.}
\end{deluxetable}

\begin{deluxetable}{lll}
\tabletypesize{\scriptsize}
\tablecaption{Clusters/Subgroups in W43 \label{clustersumW43}}
\tablewidth{0pt}
\tablehead{\colhead{Parameter} & \colhead{SLF} & \colhead{$d_c$ = 59$\arcsec$}\\
\colhead{} & \colhead{Method} & \colhead{}}\\
\startdata
Number of clusters & 51 & 95\\
Cutoff distance & 81$\arcsec$(2.20~pc) & 59$\arcsec$(1.61~pc)\\
Percent in clusters & 73 & 43\\
Group diameter & 52$\arcsec$-5412$\arcsec$ (1.4-142~pc) & 29$\arcsec$-554$\arcsec$ (0.8-14~pc)\\
Class II/I Ratio & 4.58(0.19)\tablenotemark{*} & 3.71(0.19)\tablenotemark{*}\\
\enddata
\tablenotetext{*}{Number in parentheses indicates uncertainty in ratio.}
\end{deluxetable}

\section{SED Models for MYSOs}\label{sec:sed}

We used SED models to identify the MYSO candidates in the W51 and W43 regions since it is difficult to identify them using colors. We used the Python package of the SED Fitting tool by \citet{rob07} and SED models from \citet{rob06} to model the available photometric data (2MASS/UKIDSS $J$, $H$, and $K_{s}$, \textit{Spitzer}/IRAC $3.6$, $4.5$, $5.8$, and $8.0$~$\mu$m, and \textit{Spitzer}/MIPS $24$~$\mu$m). This best-fit method is described in detail by \citet{rob07}. Only objects with at least six data points were fitted, and we used $e^{-\chi^2/2}$ values as weights. We chose the fits that satisfy the criterion of (${\chi}^2$ - ${\chi}^2 _{best}$)/$n_{data}$< 3 as in \citet{rob07}. Only objects with a minimum of 10 fits satisfying the defined best-fit criteria were used when calculating the weighted mean values for the physical parameters. We allowed the $A_{V}$ to vary between 0 and 40 mag, which gives enough flexibility and prevents unphysical solutions, and is also consistent with the $A_{V}$ values for MYSOs that are forming UC\ion{H}{2}s \citep[e.g.,][]{han02}. We also let the distance to vary within 10\% ranges of 5.4 kpc for W51, and 5.6 kpc for W43. For sources without a $24$~$\mu$m detection (except for sources in the saturated or high-background regions), we used an upper limit based on the completeness estimate at $24$~$\mu$m by \citet{gut15} in order to better constrain the SED. The luminosity and temperature are directly derived from the SED, and then mass and age values are determined using the evolutionary tracks in a Hertzprung-Russel diagram. The lack of photometric points at wavelengths longer than $24$~$\mu$m means that the disk and envelope parameters are poorly constrained. The total dust mass, for example, can only be estimated if the location of the far-IR bump is known. 

We have applied the Bayesian method described in \citet{azi15} to construct posterior probability distribution functions for the model parameters. We assume that the measurement errors in the photometry are normally distributed, which results in a likelihood function proportional to the $e^{-\chi^2/2}$ distribution. We use uniform priors for all of the model parameters. The resulting posteriors are sampled using a Markov Chain Monte Carlo (MCMC) method. Within the set of YSOs we selected the ones with $L$ $\geq$ $10^3$L$_\odot$ from both methods. We give the weighted mean values for the physical parameters such as masses and luminosities from the \citet{rob06} models, and to provide some indication of the model accuracy, we report the standard deviation values in Table~\ref{MYSOParam_W51} and Table~\ref{MYSOParam_W43}. These model grids do not include binary or multiple system cases, while the multiplicity in massive stars and in star clusters is known to be higher than that of young low-mass stars \citep{duc01,zin07}. We should also note that the Robitaille models were developed mainly for lower-mass protostars. Therefore, the physical parameters of the MYSO candidates given here should be used cautiously. However, \citet{deb16} compared the fit results from the Robitaille method to a Turbulent Core model-based method \citep{zha11} that was developed for intermediate- and high-mass stars, and they found that, although the accretion rates are significantly underestimated, the stellar masses are consistent. Moreover, \citet{mot11} found that for sources with $L$ $>$ $10^3$L$_\odot$, luminosities are better determined with the SED fitter compared to graybody fitting. However, MYSO candidates determined with luminosities over $10^3$L$_\odot$ should still be confirmed by far-IR observations and spectroscopic data.

\begin{figure}
\centering
\includegraphics[width=4cm]{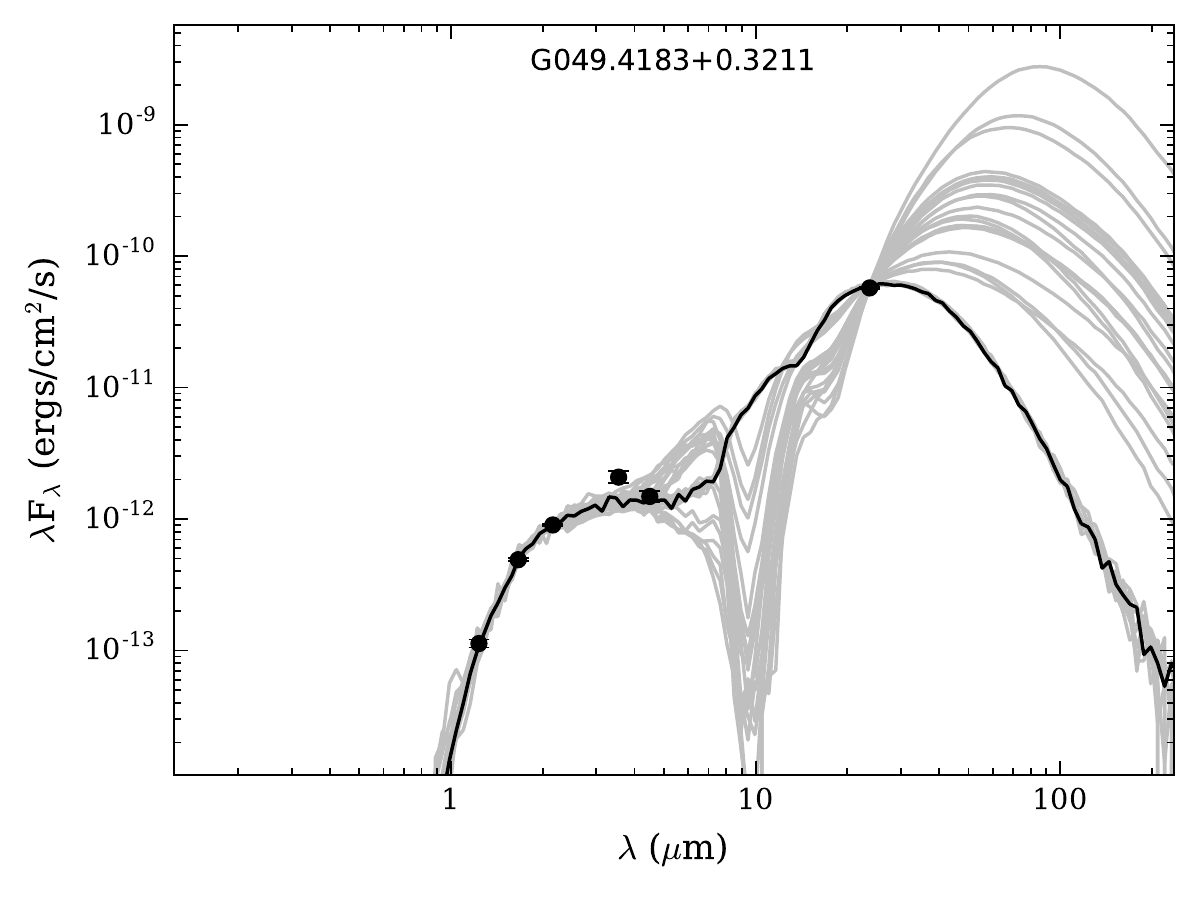}
\includegraphics[width=4cm]{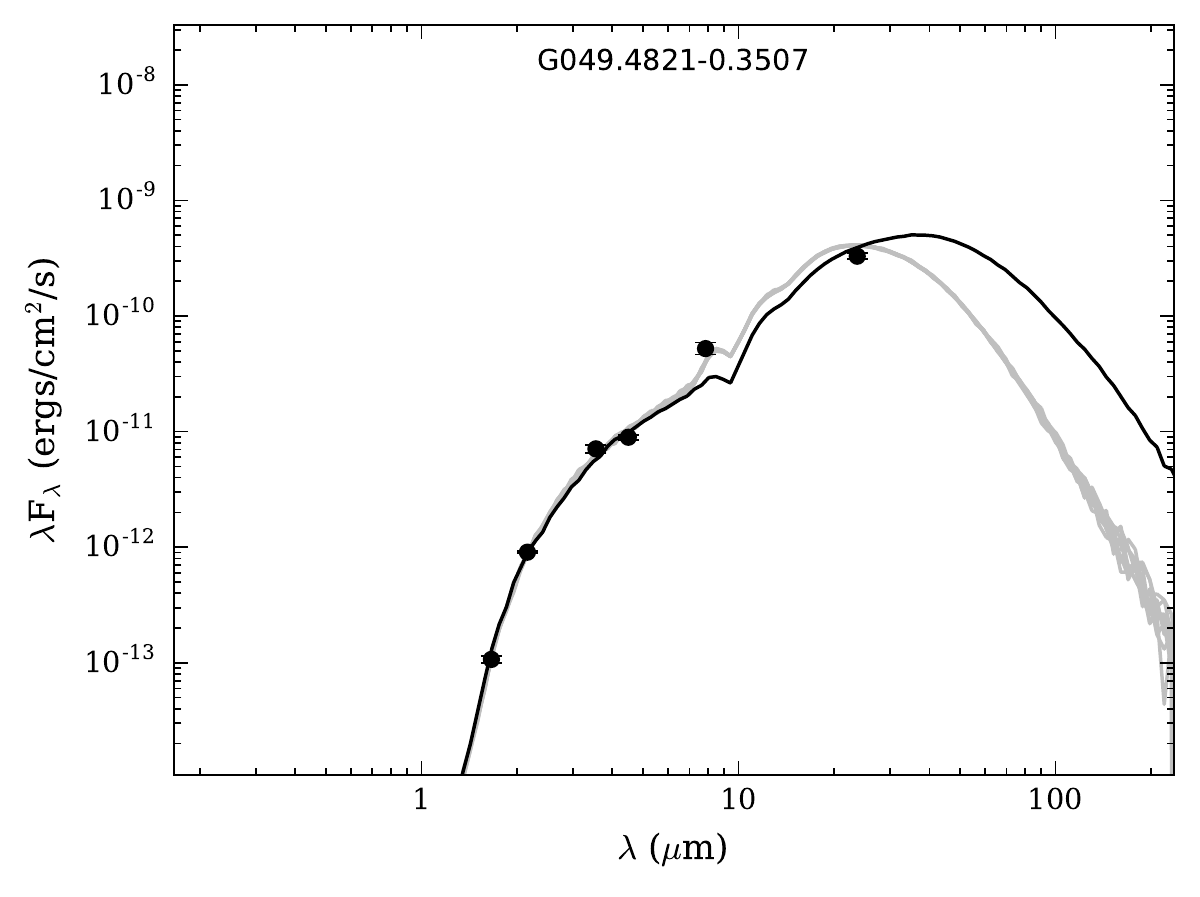}
\includegraphics[width=4cm]{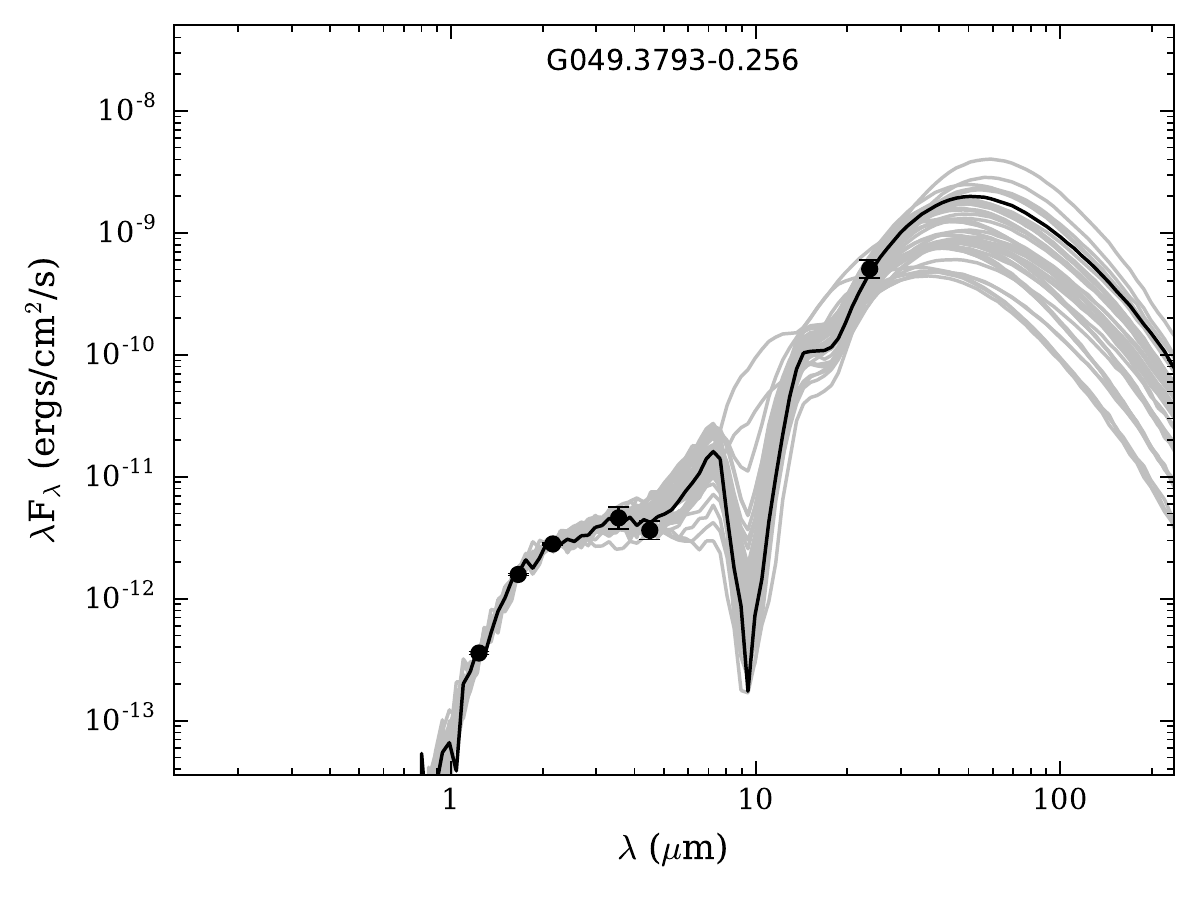}
\includegraphics[width=4cm]{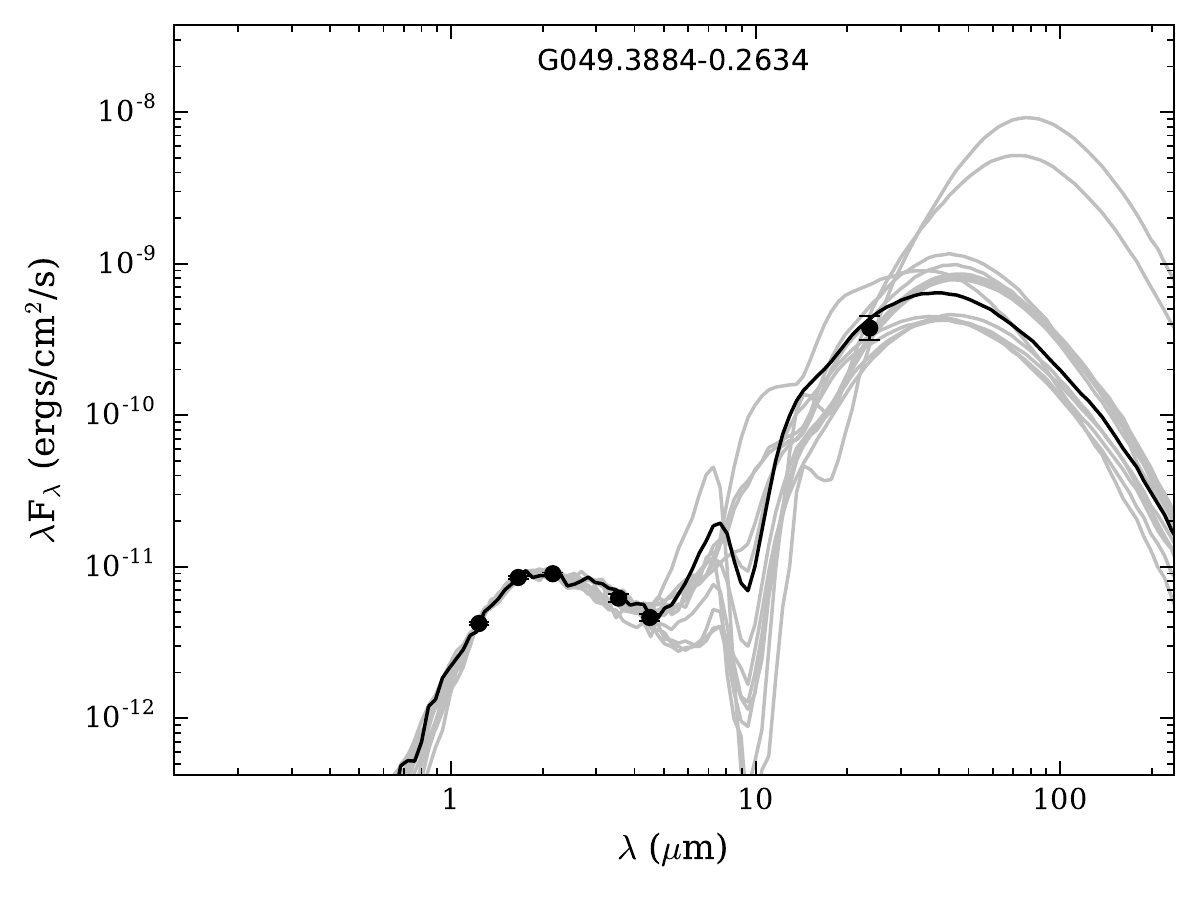}
\caption{YSO model SEDs that fit best (black line) to the data points (filled circles) in the W51 region. Gray lines show the fits satisfying (${\chi}^2$ - ${\chi}^2 _{best}$)/$n_{data}$< 3 criteria. This figure shows those sources with $24$~$\mu$m data listed in Table~\ref{MYSOParam_W51}.}\label{fig:W51_MYSO_SEDs}
\end{figure}

\begin{figure}
\centering
\includegraphics[width=4cm]{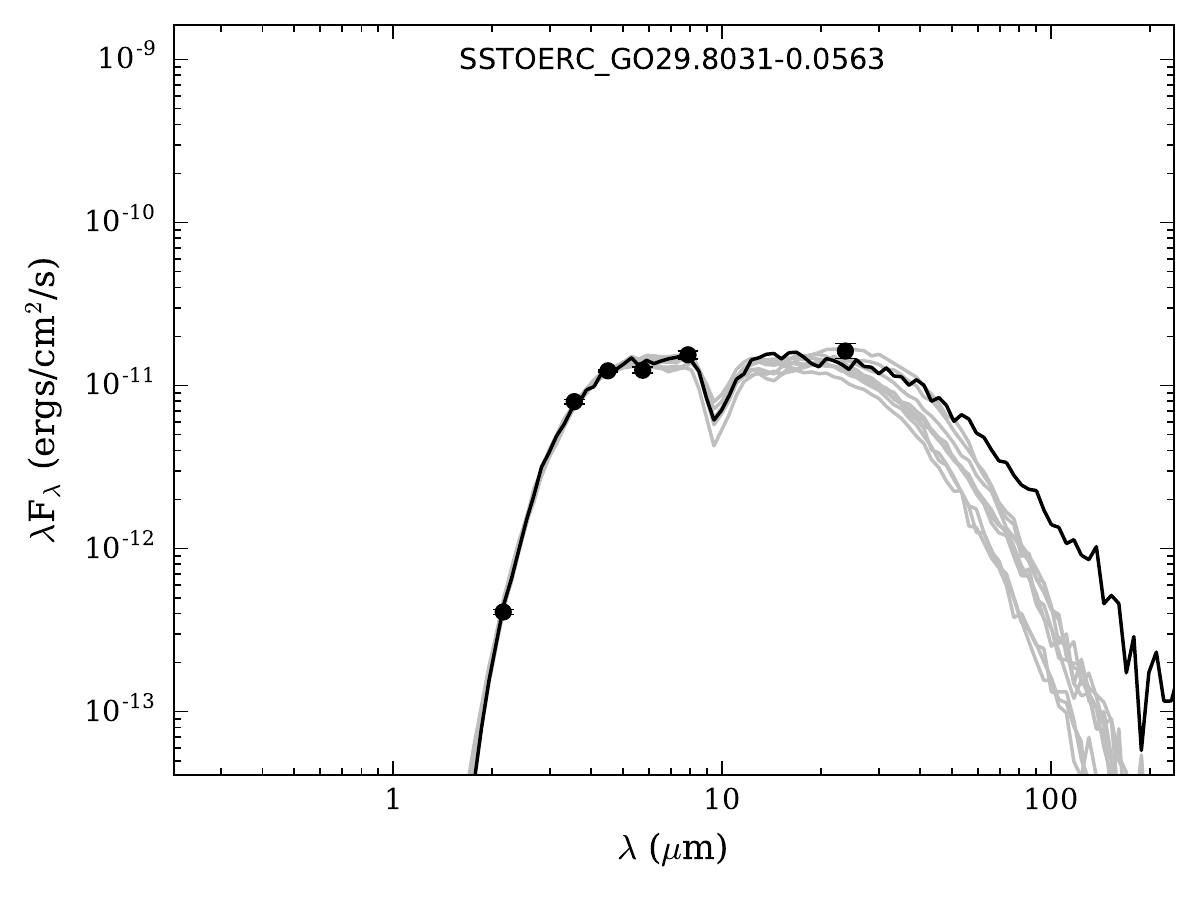}
\includegraphics[width=4cm]{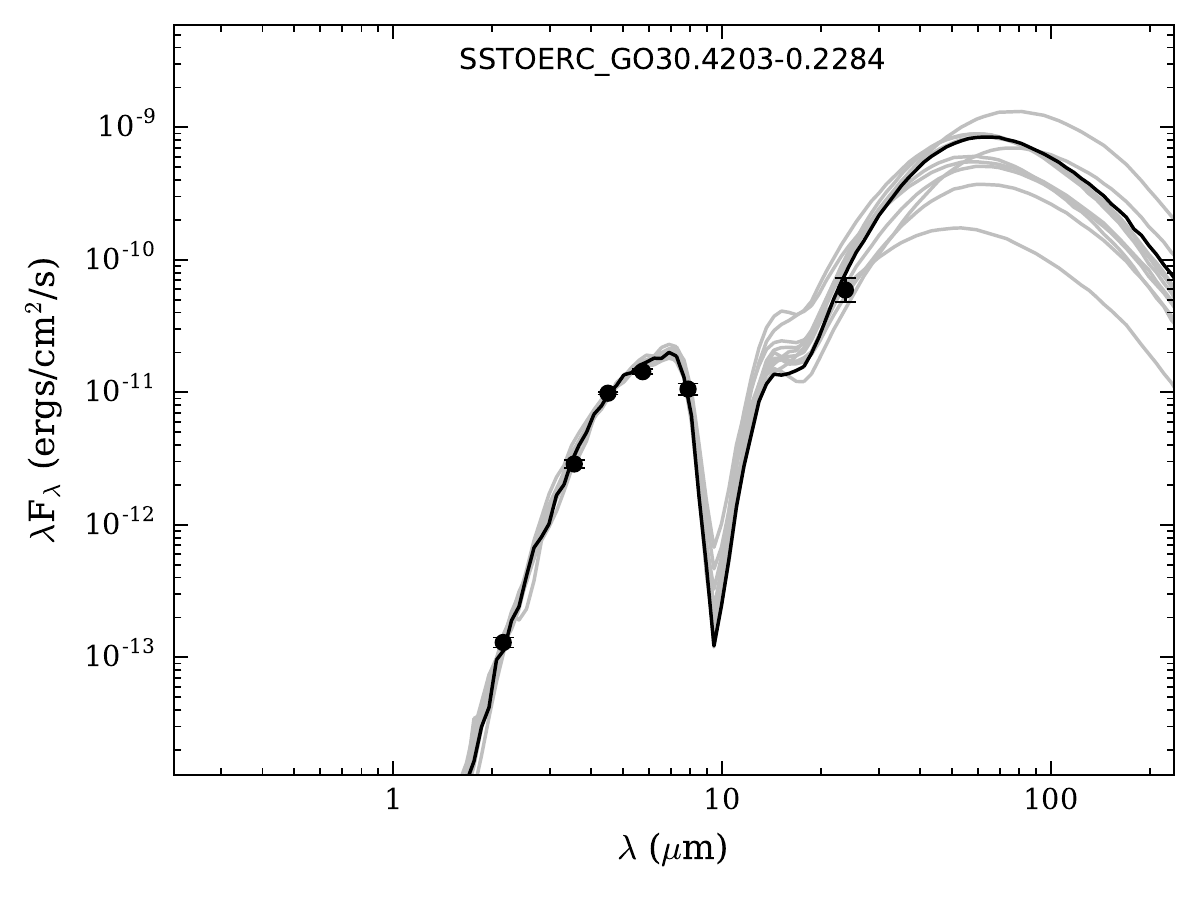}
\includegraphics[width=4cm]{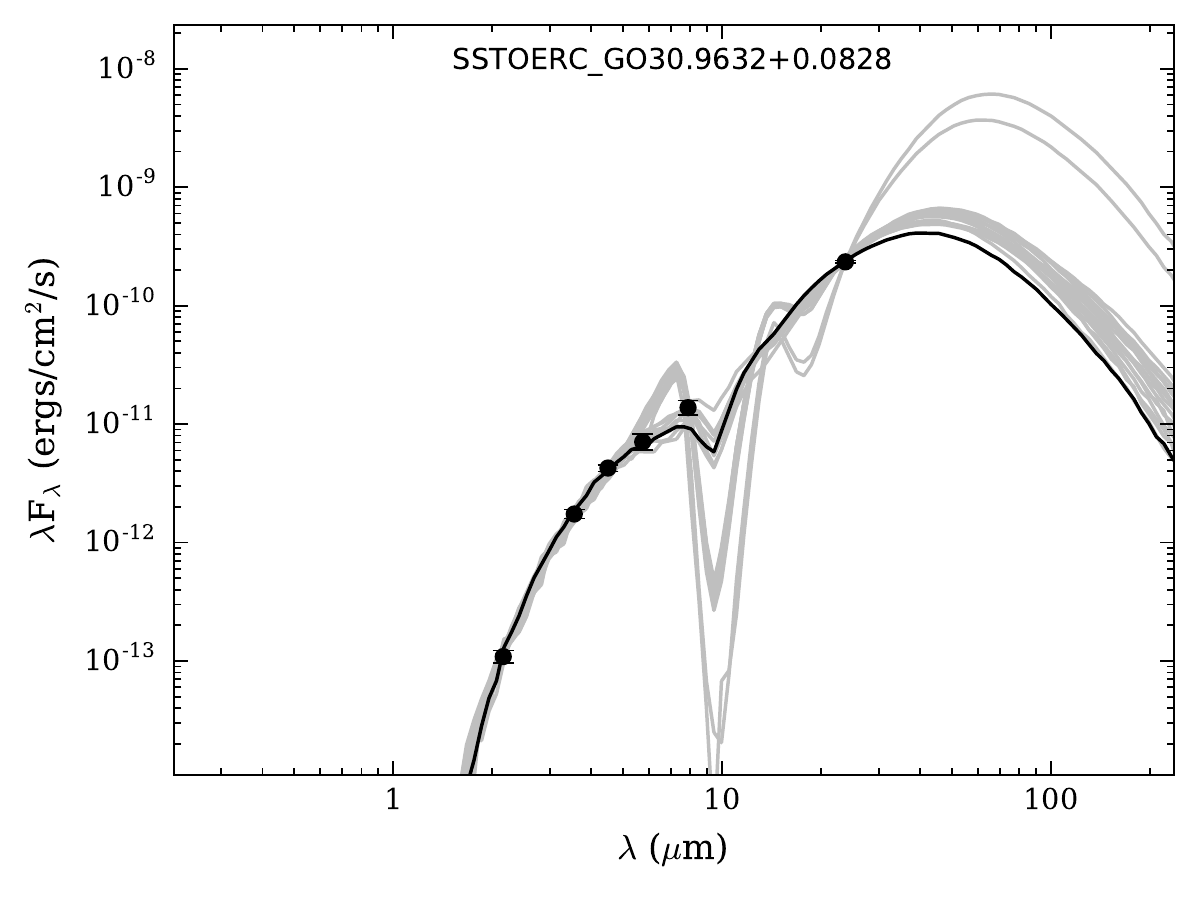}
\caption{YSO model SEDs that fit best (black line) to the data points (filled circles) in the W43 region. Gray lines show the fits satisfying (${\chi}^2$ - ${\chi}^2 _{best}$)/$n_{data}$< 3 criteria. This figure shows those sources with $24$~$\mu$m data listed in Table~\ref{MYSOParam_W43}.}\label{fig:W43_MYSO_SEDs}
\end{figure}

As a result, we identified 17 and 14 MYSO candidates in W51 and W43, respectively. In W51, 16 of the MYSO candidates are located within the clusters, and one of them is a distributed source. Seven of these MYSO candidates were classified as Class I, eight as Class II candidates, one as a transition disk candidate, and one as an embedded source candidate with IRAC-color classification. In W43, 11 of the MYSO candidates are located within the clusters and the rest are distributed outside. Among the MYSO candidates, nine were classified as Class I candidates and five were classified as Class II candidates using the IRAC-color classification. The distribution of MYSO candidates is shown in Figure~\ref{fig:W51HIIs} and Figure~\ref{fig:W43MYSOs} and the SED plots for those having $24$~$\mu$m data are shown in Figure~\ref{fig:W51_MYSO_SEDs} and Figure~\ref{fig:W43_MYSO_SEDs}.

\begin{deluxetable*}{llllcclllllllccc}
\tabletypesize{\scriptsize}
\tablecaption{Physical Parameters of Massive YSO Candidates in W51 \label{MYSOParam_W51}}
\tablewidth{0pt}
\tablehead{\colhead{No.} & \colhead{Name} &  \colhead{R.A.} & \colhead{Decl.} & \colhead{Luminosity\tablenotemark{a}} & \colhead{$\sigma$\tablenotemark{b}} & \colhead{Mass\tablenotemark{a}} & \colhead{$\sigma$\tablenotemark{b}}  & \colhead{IRAC} & \colhead{SED} & \colhead{SED}\\
\colhead{} & \colhead{} & \colhead{J2000.0($\degr$)} & \colhead{J2000.0($\degr$)} & \colhead{($10^3$L$_\odot$)} & \colhead{($10^3$L$_\odot$)} & \colhead{(M$_\odot$)} & \colhead{(M$_\odot$)}  & \colhead{Class\tablenotemark{c}} & \colhead{Class\tablenotemark{c}} & \colhead{Slope\tablenotemark{d}}}\\
\startdata
$1^{e}$ & SSTOERC$\_$G049.4183+0.3211 & 290.252258 & 14.780608 & 1.14 & 0.43 & 6.84 & 0.94 & 3 & 1 & 1.74(0.06)\\
$2^{e}$ & SSTOERC$\_$G049.4821-0.3507 & 290.895782 & 14.520227 & 4.12 & 1.58 & 8.65 & 1.27 & 1 & 1 & 2.55(0.2)\\
$3^{e}$ & SSTOERC$\_$G049.3793-0.256  & 290.759369 & 14.474311 & 2.97 & 0.99 & 10.49 & 1.22 & 0 & 1 & 2.02(0.36)\\
$4^{e}$ & SSTOERC$\_$G049.3884-0.2634 & 290.770569 & 14.478874 & 1.08 & 0.85 & 8.79 & 0.70  & 2 & F & 0.29(0.9)\\
5 & SSTOERC$\_$G048.8743-0.5188 & 290.752411 & 13.905082 & 4.06 & 1.50 & 8.72 & 1.29 & 1 & 1 & 2.12(0.12)\\
6 & SSTOERC$\_$G048.9022-0.2734 & 290.54303  & 14.045395 & 2.22 & 0.63 & 7.47 & 0.73 & 1 & 1 & 2.35(0.12)\\
7 & SSTOERC$\_$G048.9916-0.3143 & 290.623596 & 14.104997 & 5.15 & 0.12 & 9.44 & 0.09 & 2 & 1 & 1.02(0.18)\\
8 & SSTOERC$\_$G049.0693-0.3164 & 290.663391 & 14.172537 & 1.09 & 0.06 & 6.28 & 0.09 & 2 & F & 0.11(0.21)\\
9 & SSTOERC$\_$G049.1138-0.2963 & 290.666656 & 14.221247 & 1.15 & 0.0 & 6.37 & 0.00 & 2 & 1 & 0.35(0.08)\\
10 & SSTOERC$\_$G049.0327-0.3431 & 290.66983  & 14.127697 & 5.15 & 0.0 & 9.45 & 0.00  & 1 & 1 & 0.85(0.17)\\
11 & SSTOERC$\_$G049.2213-0.3354 & 290.754608 & 14.297651 & 1.17 & 0.13 & 6.39 & 0.21 & 2 & 1 & 2.02(0.27)\\
12 & SSTOERC$\_$G049.3820-0.2563 & 290.760986 & 14.476582 & 1.16 & 0.21 & 6.36 & 0.33 & 2 & 1 & 0.95(0.17)\\
13 & SSTOERC$\_$G049.3800-0.2696 & 290.772064 & 14.468561 & 5.39 & 2.36 & 8.7 & 1.47 & 1 & 1 & 2.81(0.28)\\
14 & SSTOERC$\_$G049.5688-0.2725 & 290.866852 & 14.633584 & 1.13 & 0.15 & 6.33 & 0.16 & 2 & 1 & 0.45(0.24)\\
15 & SSTOERC$\_$G049.4252-0.354  & 290.870911 & 14.468496 & 1.61 & 0.59 & 6.85 & 0.66 & 1 & 1 & 2.12(0.18)\\
16 & SSTOERC$\_$G049.4491-0.3611 & 290.889069 & 14.486297 & 1.20 & 0.31 & 6.43 & 0.30 & 2 & 1 & 0.63(0.24)\\
17 & SSTOERC$\_$G049.4885-0.3628 & 290.909912 & 14.520178 & 9.47 & 2.45 & 11.38 & 1.1 & 1 & 1 & 2.39(0.13)\\
\enddata
\tablenotetext{a}{Weighted mean results from the SED Fitting tool by \citet{rob07}.}
\tablenotetext{b}{Standard deviation of the weighted mean values are reported here. ${\chi}^2$ values are used to weight the fits. Minimum 10 fits from the defined best fit criteria was required to calculate the weighted mean values.}
\tablenotetext{c}{0: deeply embedded, 1: Class I, 2: Class II, F: Flat spectrum, 3: transition disks.}
\tablenotetext{d}{ Values in parentheses signify error in last two digits of the SED slope value.}
\tablenotetext{e}{Sources with $24$~$\mu$m data.}
\end{deluxetable*}

\begin{deluxetable*}{llllccccccl}
\tabletypesize{\scriptsize}
\tablecaption{Physical Parameters of Massive YSO Candidates in W43 \label{MYSOParam_W43}}
\tablewidth{0pt}
\tablehead{\colhead{No.} & \colhead{Name} &  \colhead{R.A.} & \colhead{Decl.} & \colhead{Luminosity\tablenotemark{a}}  & \colhead{$\sigma$\tablenotemark{b}} & \colhead{Mass\tablenotemark{a}}& \colhead{$\sigma$\tablenotemark{b}}  & \colhead{IRAC} & \colhead{SED} & \colhead{SED}\\
\colhead{} & \colhead{} & \colhead{J2000.0($\degr$)} & \colhead{J2000.0($\degr$)} & \colhead{($10^3$L$_\odot$)} & \colhead{($10^3$L$_\odot$)} & \colhead{(M$_\odot$)} & \colhead{(M$_\odot$)} & \colhead{Class\tablenotemark{c}} & \colhead{Class\tablenotemark{c}} & \colhead{Slope\tablenotemark{d}}}\\
\startdata
$1^{e}$  & SSTOERC$\_$G029.8031-0.0563 & 281.481842 & -2.810057 & 4.58 & 3.62 & 8.72 & 2.92  & 1 & 1 & 2.42(1.04)\\
$2^{e}$  & SSTOERC$\_$G030.4203-0.2284 & 281.917053 & -2.339325 & 1.52 & 0.51 & 8.24 & 0.46  & 1 & 1 & 2.92(1.14)\\
$3^{e}$   & SSTOERC$\_$G030.9632+0.0828 & 281.887787 & -1.714331 & 2.24 & 0.88 & 7.29 & 0.75  & 1 & 1 & 2.67(0.19)\\
4  & SSTOERC$\_$G030.9238-0.1319 & 282.060944 & -1.847242 & 1.43 & 1.45  & 6.48 & 1.24 & 1 & 1 & 1.25(0.45)\\
5  & SSTOERC$\_$G030.5565-0.0894 & 281.855469 & -2.154748 & 1.15 & 0.03 & 6.37 & 0.06 & 1 & F & 0.28(0.56)\\
6  & SSTOERC$\_$G030.7019-0.0338 & 281.872314 & -2.000047 & 1.44 & 0.94 & 6.58 & 0.99  & 1 & 1 & 0.64(0.22)\\
7  & SSTOERC$\_$G030.6487-0.2003 & 281.996246 & -2.123312 & 1.02 & 0.35 & 6.17 & 0.85  & 1 & 1 & 2.10(0.31)\\
8  & SSTOERC$\_$G029.2101+0.1301 & 281.044617 & -3.25246  & 1.46 & 1.12 & 6.52 & 1.24  & 1 & 1 & 1.48(0.34)\\
9  & SSTOERC$\_$G030.0299+0.308  & 281.260956 & -2.441933 & 1.04 & 0.02 & 6.21 & 0.05  & 1 & 1 & 0.37(0.13)\\
10 & SSTOERC$\_$G031.4202-0.5835 & 282.689545 & -1.61143  & 1.20 & 1.25 & 5.37 & 2.3   & 2 & 1 & 1.06(0.36)\\
11 & SSTOERC$\_$G029.6445-0.1788 & 281.518433 & -3.00707  & 1.09 & 0.25 & 6.24 & 0.48  & 2 & F & 0.23(0.28)\\
12 & SSTOERC$\_$G030.6549-0.1449 & 281.949799 & -2.092496 & 2.08 & 2.64 & 6.23 & 2.81  & 2 & 1 & 1.25(0.39)\\
13 & SSTOERC$\_$G030.5984-0.1106 & 281.893402 & -2.127159 & 1.61 & 0.05 & 6.95 & 0.06  & 2 & 1 & 1.71(0.3)\\
14 & SSTOERC$\_$G029.8775-0.0379 & 281.49942  & -2.735451 & 1.15 & 0.01 & 6.37 & 0.02  & 2 & F & -0.11(0.34)\\
\enddata
\tablenotetext{a}{Weighted mean results from the SED Fitting tool by \citet{rob07}.}
\tablenotetext{b}{Standard deviation of the weighted mean values are reported here. ${\chi}^2$ values are used to weight the fits. A minimum of 10 fits from the defined best-fit criteria was required to calculate the weighted mean values.}
\tablenotetext{c}{0: deeply embedded, 1: Class I, 2: Class II, F: Flat spectrum, 3: transition disks.}
\tablenotetext{d}{ Values in parentheses signify error in last two digits of the SED slope value.}
\tablenotetext{e}{Sources with $24$~$\mu$m data.}
\end{deluxetable*}
 
\section{Discussion}\label{sec:discuss}

\subsection{Massive Star Formation Tracers}

\subsubsection{UC\ion{H}{2} Regions and Masers in W51}

\begin{figure*}
\centering
\includegraphics[width=14cm]{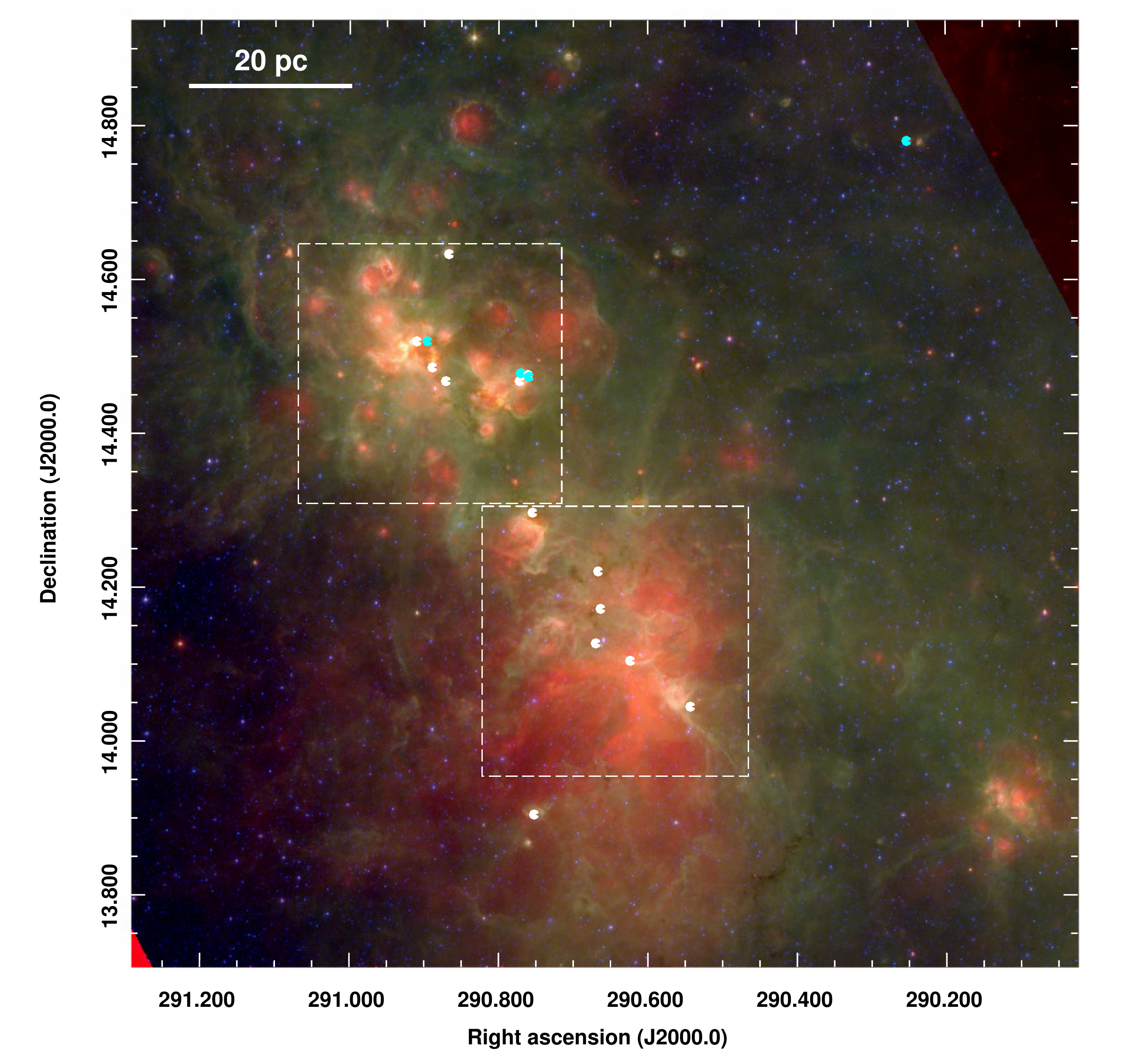}
\includegraphics[width=16cm]{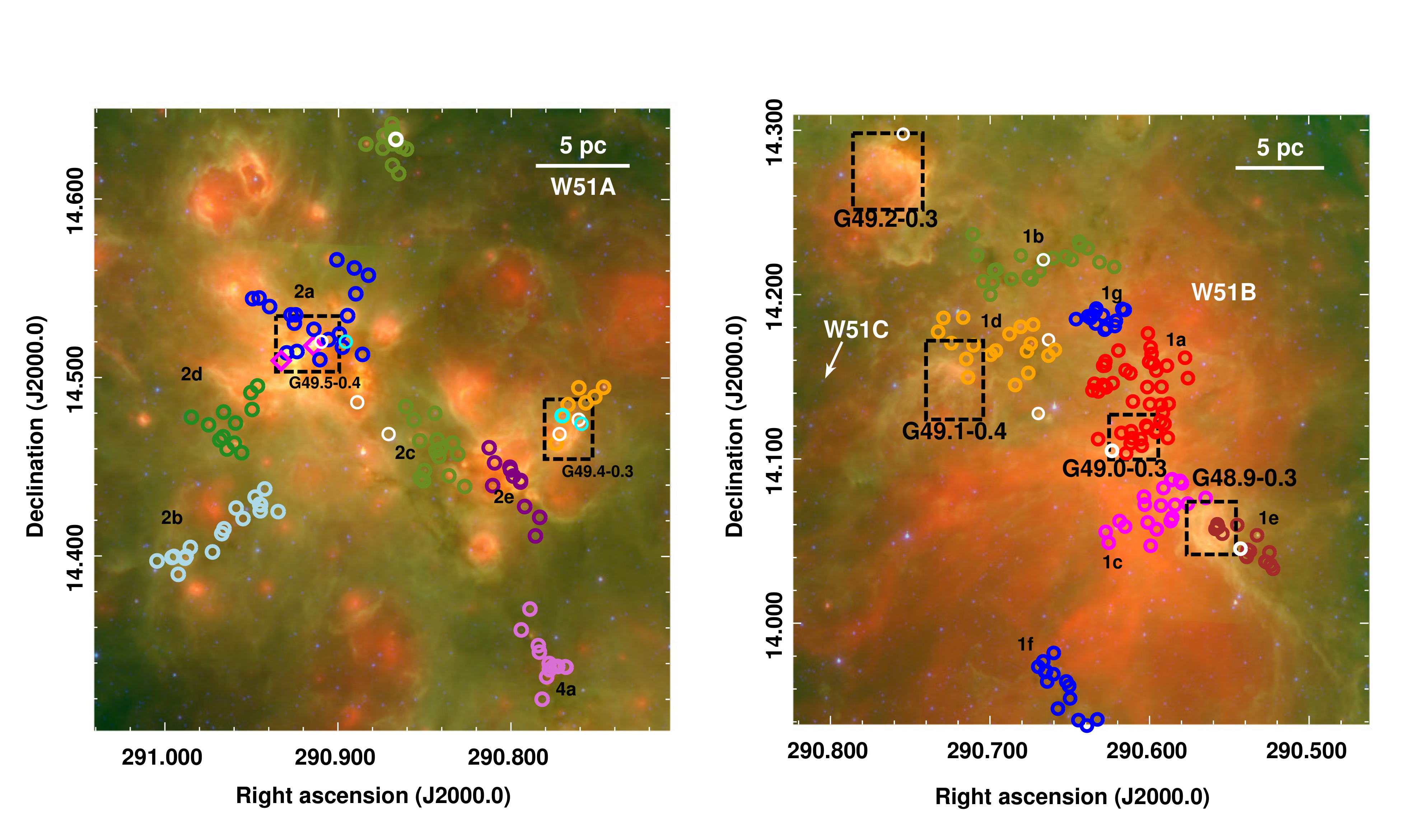}
\caption{Top: the distribution of MYSO candidates is shown in the RGB image (blue: 3.6 $\mu$m, green: 8.0 $\mu$m, red: 24.0 $\mu$m). White circles represent the MYSO candidates, and cyan circles represent the MYSO candidates with 24.0 $\mu$m detections. Bottom: RGB images of W51A (as shown by the dashed box in the upper panel) and W51B (as shown by the dashed box in the upper panel). Magenta diamonds show methanol masers, and black dashed boxes show the main \ion{H}{2} regions. Subgroups of YSOs are shown with different colored circles, and MYSO candidates are shown with white circles. YSO candidates that do not belong to subgroups are not shown here.}\label{fig:W51HIIs}
\end{figure*}

W51 hosts several \ion{H}{2} regions: G49.5-0.4 and G49.4-0.3 associated with W51A, and G49.2-0.3, G49.1-0.4, G49.0-0.3, and G48.9-0.3 associated with W51B \citep{kum04} (Figure~\ref{fig:W51HIIs}). In addition, there are several compact and UC\ion{H}{2} regions that are not detected at <20$\mu$m because of heavy extinction \citep{gau93,zha95,zha97}. 

W51A, the most active part of the region, hosts several protoclusters and UC\ion{H}{2} regions \citep{gau93,gol94,zha95,bar08}. G49.5-0.4 is the most luminous part in W51A and hosts two embedded protoclusters strongly correlated with radio continuum emission: IRS 1 and IRS 2 \citep{woo89,wyn94}. In the IRS 2 UC\ion{H}{2} region, a stellar object that is not resolved in our images was identified by \citet{fig08} and confirmed as an early O-type object with spectroscopy by \citet{bar08}. We identified one subgroup (2a in Table~\ref{W51GMCclusters}) that covers both protoclusters including the \ion{H}{2} region G49.5-0.4 as seen in Figure~\ref{fig:W51HIIs}. Half of the YSO candidates of subgroup 2a are positionally associated with the G49.5-0.4, and the other half surrounds the \ion{H}{2} region with a filamentary shape. We identified two MYSO candidates, SSTOERC$\_$G049.4885-0.3628 and SSTOERC$\_$G049.4821-0.3507 (detected also at $24\mu$m), that are shown in Figure~\ref{fig:W51HIIs}. The other main \ion{H}{2} region in W51A is G49.4-0.3, which seems to be associated with subgroup 2f, where we identified 11 YSO candidates, four of which are MYSO candidates: SSTOERC$\_$G049.3820-0.2563, SSTOERC$\_$G049.3800-0.2696, SSTOERC$\_$G049.3884-0.2634 and SSTOERC$\_$G049.3793-0.256.

In W51B, the largest and brightest \ion{H}{2} region is G49.2-0.3, which is indicated as an interacting region with the W51C supernova remnant \citep{gre97,koo97}. We identified a MYSO candidate in the vicinity of this region, SSTOERC$\_$GO49.2213-0.3354, and five more YSO candidates. The other three \ion{H}{2} regions, G49.1-0.4, G49.0-0.3, and G48.9-0.3, seem positionally but partially associated with subgroups 1d, 1a, and 1e, respectively, and are located along a filamentary structure parallel to the Galactic plane (right panel of Figure~\ref{fig:W51HIIs}). We identified one MYSO candidate, SSTOERC$\_$G048.9916-0.3143, in the field of G49.0-0.3 \ion{H}{2} region, which is a weaker \ion{H}{2} region in W51B. And finally, G48.9-0.3, a strong UC\ion{H}{2} region in W51B, is partially corresponds to the subgroup 1e and is also located at the edge of the subgroup 1c, which hosts the MYSO candidate, SSTOERC$\_$GO48.9022-0.2734 in its vicinity. 

W51 also hosts many masers \citep{zha95,ima02,fis07,hen13}. Among them are methanol masers, which are known as good massive star formation tracers \citep{pes02,bre13}. Two main methanol maser regions toward W51A \citep[G049.49-00.37 and G49.49-0.39;][]{cas95,min00,eto12} are shown in Figure~\ref{fig:W51HIIs}. There are tens of methanol masers in that region clustered within a few arcsec, and most of them do not have known counterparts with any IR source. This indicates that these sources are in very early evolutionary stages that we cannot yet trace  with IR photometry. 

\subsubsection{Compact Fragments and Masers in W43}

\begin{figure*}
\centering
\includegraphics[width=18cm]{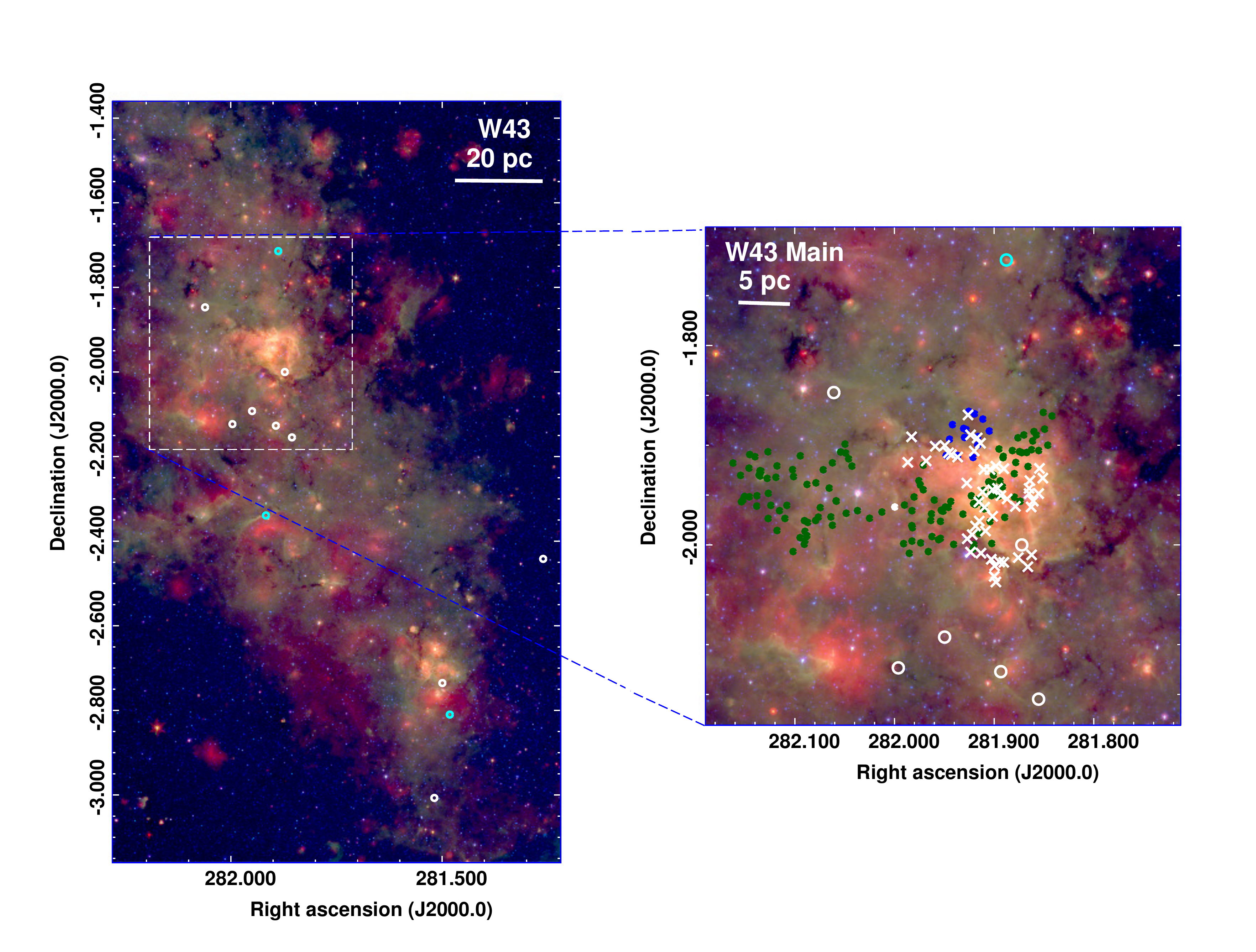}
\caption{Distribution of MYSO candidates in W43 is shown in the left RGB image. The distribution is shown in the giant \ion{H}{2} region in W43-Main on the right, where \citet{mot03} studied the compact sub-millimeter fragments that are shown with white Xs (blue: 3.6 $\mu$m, green: 8.0 $\mu$m, red: 24.0 $\mu$m). Subgroup 3 and subgroup 62 are shown in order to illustrate the positional correspondence with the sub-millimeter fragments. White circles represent the MYSO candidates, and cyan circles represent the MYSO candidates with 24.0 $\mu$m detections.}\label{fig:W43MYSOs}
\end{figure*}

W43 hosts a giant \ion{H}{2} region dominated by its centimeter and infrared emission, which makes the detection of compact sources with associated UC\ion{H}{2}s difficult \citep{mot03}. Using 1.3 mm and 350 $\mu$m continuum emission, \citet{mot03} identified 50 compact fragments in the W43-Main region with sizes ranging from 0.09 to 0.56 pc, which tens of fragments of which are protoclusters \citep[see][]{lou14}. \citet{mot03} suggested that the fragments MM1, MM2, and MM3, associated with masers and outflows, are forming high-mass stars;  \citet{bal10} confirmed this with High-GAL far-infrared observations. These sources should be in very early stages (Class 0-like) and be still very embedded. Consequently, current observations have not yet found many associated near infrared sources within the fragments. Recently, 20 high-mass protostars were found in the MM1 ridge, encompassing MM1 and MM9 (Motte et al. 2017, in preparation). We did not find any MYSO candidates associated with these fragments. However, in subgroup 3 we see a YSO candidate coincident with each of the following fragments: MM4, MM11, MM14, MM15, and MM20.  \citet{mot03} indicates that MM11 is also a methanol source with no infrared or centimeter detection; however, we see a Class I candidate (SSTOERC$\_$GO30.7598$-$0.0525) that coincides with the filament. We also see a YSO candidate positionally associated with the fragments MM9 and MM16 fragments that corresponds to subgroup 62 (Figure~\ref{fig:W43MYSOs}), which could become more massive in the future given the large amount of gas surrounding it.

\subsection{Star Formation History in W51}

In Section \ref{sec:clustering}, we identified YSO groups and subgroups by using the MST method. In order to understand the star formation history in the region, we should consider the relative ages of these groups. Assuming Class I sources  represent relatively early stages of formation compared to Class II sources, we can use the ratio of Class II/I sources to compare the relative ages of different groups within the region. 

The Class II/I ratio is $2.50\pm0.25$ when we consider all of the MST groups identified with $d_c$ = 82$\arcsec$.5(2.17~pc; see Table~\ref{clustersumW51}). If we only consider the subgroups identified with $d_c$ = 51$\arcsec$, we find a Class II/I ratio of  $2.06\pm(0.27)$, which is consistent with the ratio from the SLF method within the uncertainties. If we look at individual groups within the region, the biggest group, G49.01-0.31, consists of seven subgroups (the red group in Figure~\ref{fig:W51mstresults2}, corresponds to W51B) and the ratio of Class II/I is $2.98\pm0.48$; if we consider only the subgroups, it is $2.56\pm0.54$. Meanwhile, for the second largest group, G49.45-0.37 (the blue group in Figure~\ref{fig:W51mstresults2}, corresponds to W51A), the ratio of Class II/I is $1.94\pm0.32$, and it consists of six subgroups with a Class II/I ratio of $1.65\pm0.44$. Therefore, either considering the large MST groups or subgroups,  G49.45-0.37 (where W51A is located) is relatively younger than G49.01-0.31 (where W51B is located). This is consistent with the \citet{gin15} study that showed a low star formation potential of W51B, with a density of $n\sim1\times10^4$ \ensuremath{\textrm{cm}^\textrm{-3}}. Furthermore, they associated the high-mass density, n $\gtrsim$ $5\times10^5$  \ensuremath{\textrm{cm}^\textrm{-3}} in W51A with high-mass star formation.
 
\citet{kum04} suggested that the star formation in W51 is triggered by galactic spiral density waves. Also, the interaction between the \ion{H}{2} region G49.2-0.3 and the W51C supernova remnant W51C \citep[$\sim3\times10^4$ years old;][]{koo95} might be triggering a future star formation episode \citep{gre97, koo97}. Furthermore, the existence of the supernova remnant and old, diffuse \ion{H}{2} regions in W51B can be considered as signposts of older star-forming regions. We did not identify a subgroup corresponding to the \ion{H}{2} region G49.2-0.3, but we identified the larger-scale cluster 4 with a Class II/I ratio of 1.70(0.39), which indicates a relatively younger age than W51B that is consistent with a triggered star formation scenario. 

\subsection{Star Formation History in W43}

The class II/I ratio is found to be $4.58\pm0.19$ for the W43 region when we consider all of the MST groups identified with $d_c$$ = $81$\arcsec$ (Table~\ref{clustersumW43}). If we only consider the subgroups, we find a smaller ratio, $3.71\pm0.19$, but it is still closer to the value from the SLF method. For just MST group 1 (the big red group in Figure~\ref{fig:MSTclustersW43}), corresponding to the whole molecular cloud, the II/I ratio is $4.20\pm0.20$. We see that group 1, identified using SLF method, consists of 62 subgroups with a Class II/I ratio of $4.19\pm0.25$. For both the SLF method and the maximum number of subgroups, W43 has significantly larger class II/I ratio than W51. The findings of many YSOs and subgroups are also consisted with the substructures found by \citet{car13} and supporting the hypothesis of several HI-to-H2 transition layers that had been suggested by \citet{bia17}.

\subsection{Comparison of Other Star-Forming Regions}

W51, W43, and W49 are among the brightest star-forming regions in the Galaxy. Although they have similar gas masses and stellar masses (see Section \ref{sec:int}), they have different morphologies and clustering properties that suggest a different star-formation history. W49 shows a hierarchical network of filaments at scales from $\sim$10 to 100 pc, suggesting that it is formed from global gravitational contraction \citep{mad13}; it also hosts a large central cluster with tens of massive YSO candidates which might be triggering the star formation. On the other hand, both W51 and W43 show a younger and an older part, with more distributed clusters and distributed MYSO candidates within the cloud.

W51 has an extended structure that shows independent sites of star formation \citep{cla09}; on the other hand, \citet{kum04} suggested that the star formation in W51 is triggered by galactic spiral density waves. \citet{cla09} derived an age range of 3-6 Myr (assuming the distance as 6 kpc) for a very luminous supergiant, suggesting at least 5 Myr of massive star formation in W51. On the other hand, by studying the old Wolf-Rayet stars \citet{bik05} determined an age range of 5-10 Myr. We have identified one large cluster, representing both W51A and W51B regions in the GMC with different Class II/I ratios, that suggests a relative age difference. In addition, the interaction between the \ion{H}{2} region G49.2-0.3 and the W51C supernova remnant \citep{gre97, koo97} might trigger future star formation. Similarly, W43 extends over 140 pc and is surrounded by an atomic gas envelope with a diameter of 290 pc (see Section \ref{sec:int}). Considering the similar distance of both regions, the class II/I ratios show that W51 is relatively younger, with the Class II/I ratio of 2.50 (0.25) than W43 which has a Class II/I ratio of 4.58(0.19). However, we should note that although the number of Class IIs are over-numerous in W43, there is a larger number of Class 0s and even Class Is which we cannot probe at near-IR. 

Among these three regions, W49 might be the youngest, with a suggested age range of 1-2 Myr \citep{hom05}. W51 is also relatively young, with a Class II/I ratio of 2.50. Even though W49 has a smaller Class II/I ratio, since W49 is a much more distant region it does not allow us to make a direct comparison as we discussed in \citet{sar15}. In addition, we should note that several SF events are developing in these clouds. For instance, W43-Main could also be at an earlier stage than W49 since it has a very compact massive cluster, formed by the present starburst cluster, in its center. This compact massive cluster should be observed by the James Webb Space Telescope in order to fully resolve and detect the individual sources in the cluster.

In addition, we see that the fraction of YSOs in clusters is slightly higher in W43 (43-73\%) than in W51 (32-60\%). Clustered fractions were reported to be 53-67\% in the Cygnus-X \citep{bee10}, 44-70\% in W5 \citep{koe08}, 53-67\% and 44-76\% in G305 and 54-81\% in G333, respectively \citep{sar15}. Although we can say that all of these regions have similar clustered fractions for all masses of YSOs, the clustered fractions for MYSOs are higher (94\% in W51, 79\% in W43, 100\% in W49) which supports the theory that massive stars are being formed in clusters.

When we compare the branch length distances shown in Table~\ref{clustersum}, we can see that both W51 (82$\arcsec$.5(2.17 pc) and W43 (81$\arcsec$(2.20 pc) have similar branch lengths, suggesting that the density of clusters is the same. These differ from G305 and also G333 \citep{sar15}, which have smaller branch length distances, indicating denser clusters than those of W51, W43, and W49. W49 has the largest branch length; however, this is likely due to its larger distance and incompleteness effects. For example, even though W49 might be intrinsically denser than the W51 and W43 regions, it has a larger branch length (96$\arcsec$ or 5.2 pc) than W51 and W43, which indicates that the clusters in W49 are less dense. However, this might be due to the larger distance and incompleteness effects. In order to make a direct comparison, one needs to rescale the source positions and magnitudes in order to determine the projected appearances of these regions at 11 kpc.

As we did in \citet{sar15} for G305 and G333, by following the same strategy we determined the projected appearance of the W51 and W43 YSOs by rescaling the relative source positions and magnitudes for a distance shift from 5.4 kpc (W51) and 5.6 kpc (W43) to the 11 kpc distance of W49. Following the rescaling, we performed the source classification and clustering as described in Section \ref{sec:yso} and Section \ref{sec:clustering}. At reprojected distances we identified 458 and 1524 YSO candidates in W51 and W43, respectively, and branch lengths of $d_c$ = 120$\arcsec$.94 (6.47 pc) with five clusters in W51, and $d_c$ = 91$\arcsec$.82 (4.91 pc) with eight clusters in W43. As seen from the clustering results for the reprojected regions, the clustering analysis yields higher branch lengths due to larger distances and incompleteness effects. The larger derived branch length is a result of our inability to resolve the smaller distances between the closest sources. The rescaled W51, which has the largest branch length of $d_c$ = 120$\arcsec$.94 (6.47 pc), seems less dense than other regions. The rescaled W43 region, with a branch length of $d_c$ = 91$\arcsec$.82 (4.91 pc), is denser than both W49 and rescaled W51 but less denser than G305 and G333. Among these regions, G305 has the densest clusters, with a branch length of $d_c$ = 52$\arcsec$ (0.90 pc). The branch length and clustered fraction values are summarized in Table~\ref{clustersum}.

\begin{deluxetable*}{lclclc}
\tabletypesize{\scriptsize}
\tablecaption{Clustering Analysis Summary \label{clustersum}}
\tablewidth{0pt}
\tablehead{\colhead{Region} & Number of Clusters & \colhead{Cutoff} & \colhead{Clustered} & \colhead{Class II/I} & Method\\
\colhead{} & \colhead{} & \colhead{distance ($\arcsec$)} & \colhead{Fraction} & \colhead{Ratio} &\colhead{}} 
\startdata
W51 & 9 & 82$\arcsec$.5(2.17 pc) & 60 & 2.50(0.25) &  \tablenotemark{a}\\
	& 16 & 51$\arcsec$(1.34 pc) & 32 & 2.06(0.27) & \tablenotemark{b}\\
W43 & 51 & 81$\arcsec$(2.20 pc) & 73 & 4.58(0.19) & \tablenotemark{a} \\
    & 95 & 59$\arcsec$(1.61 pc) & 43 & 3.71(0.19) & \tablenotemark{b}\\
G305\tablenotemark{c} & 15 & 48$\arcsec$(0.90 pc) & 76 & 3.50(0.30) & \tablenotemark{a}\\
     & 32 & 30$\arcsec$(0.60 pc) & 44 & 3.00(0.30) & \tablenotemark{b}\\
G333\tablenotemark{c} & 11 & 73$\arcsec$(1.30 pc) & 81 & 4.50(0.40) & \tablenotemark{a}\\
    & 45 & 52$\arcsec$(0.90 pc) & 54 & 4.00(0.40) & \tablenotemark{b}\\
W49\tablenotemark{c} & 3 & 96$\arcsec$(5.20 pc) & 52 & 2.11(0.43) & \tablenotemark{a} \\
    & 7 & 40$\arcsec$(2.20 pc) & 21 & 2.00(0.64) & \tablenotemark{b}\\
\hline
\hline
\\[\dimexpr-\normalbaselineskip+4pt]
Rescaled W51 & 5 & 120$\arcsec$.94(6.47) & 74 & 1.61(0.21) & \tablenotemark{a} \\
Rescaled W43 & 8 & 91$\arcsec$.82(4.91 pc) & 73 & 2.79(0.21) & \tablenotemark{a} \\
Rescaled G305\tablenotemark{c} & 4 & 52$\arcsec$(2.78 pc) & 73 & 0.55(0.15) & \tablenotemark{a} \\
Rescaled G333\tablenotemark{c} & 2 & 78$\arcsec$(4.22 pc) & 70 & 0.03(0.04) & \tablenotemark{a} \\
\enddata
\tablenotetext{a}{Straight-line fit method}
\tablenotetext{b}{The method using the short branch length distance enables us to see subgroups in the region.}
\tablenotetext{c}{\citet{sar15}}
\end{deluxetable*}

\section{Summary}\label{sec:sum}

In this paper, we generated photometric catalogs for W51 and W43 with a wavelength range of 1.2 to 24~$\mu$m by combining 2MASS/UKIDSS, Spitzer IRAC, and MIPS data in order to identify and classify the YSO candidates. We identified 302 Class I candidates, 1178 Class II/transition disk candidates, and 56 embedded sources in W51, and 917 Class I candidates, 5187 Class II/transition disk candidates, and 144 embedded sources in W43. In addition, we compared the color classification method to the SED slope method and concluded that except for a small percentage of sources, both methods are consistent. Following the YSO classification, we identified groups and clusters based on their spatial distributions using the MST method. We identified nine clusters in W51, where 60\% of the YSO candidates (including transition disks) belong to the clusters, each with a minimum of 10 members. We identified 51 clusters in W43 (over six times more than in W51), where 73\% of the YSO candidates belong to the clusters. Similar branch length distances ($\sim$2 pc) in W51 and W43 indicate similar densities within clusters in both regions. However, with a larger Class II/I ratio, W43 (4.58) is a relatively older region than W51 (2.50), and W43 has at least six times more clusters. Based on the morphology of the clusters, which are distributed within the cloud and show relative age differences according to their Class II/I ratios and also the existence of Class 0 clusters studied in the FIR and sub-millimeter, there might be several sites of star formation that are independent of one another in terms of their ages, physical conditions, and evolutionary progress. We performed 1000 Monte Carlo simulations of randomly distributed objects in order to compare the properties of random clusters to real ones, and also applied a K$-$S test in order to test the statistical significance of the identified clusters. We find that for both regions there is a high probability that the identified MST groups are different than expected from a random distribution.

We identified 17 and 14 MYSO candidates ($L$ $\geq$ $10^3$L$_\odot$) in W51 and W43, respectively, according to both the SED Fitting tool by \citet{rob07} and the Bayesian method by \citet{azi15}. The clustered fraction of MYSOs are $\sim$94\% and $\sim$79\% in W51 and W43, respectively. In both W51A and W51B, we see subgroups with MYSO candidates associated with \ion{H}{2} regions, which makes these sources strong candidates for follow-up spectroscopic observations. We do not see any of our MYSO candidates associated with previously identified dense fragments in W43, which might be a sign of several independent star formation events and bursts, while we see YSO candidates in subgroup 3 that are associated with dense fragments. These sources are good candidates for follow-up studies. 

We compared W51 and W43 to each other and also to other star-forming regions such as G305, G333, and W49. While W51 is a smaller star-forming region than W43, W51 and W43 have many similarities, such as distributed clusters in the younger and older parts of each cloud, suggesting independent star formation throughout the clouds. Both W51 and W43 have similar YSO densities in clusters, while G305 and G333 have smaller branch length distances, indicating denser clusters. Compared to the others discussed in this study and W49, G305 and G333 are small star-forming regions \citep{sar15}. W49 is smaller than both W51 and W43 and has more centrally concentrated cluster formation. 

\acknowledgments
This work is based in part on observations made with the {\it Spitzer Space Telescope}, which is operated by the Jet Propulsion Laboratory, California Institute of Technology, under a contract with NASA. Support for this work was provided by NASA. G.S. acknowledges support from NASA grant NNX12AI60G, Istanbul University grant BAP50195, Marie Sklodowska-Curie RISE grant "ASTROSTAT" (project number 691164), and Swiss National Science Foundation Sinergia project "STARFORM" (project number CH: CRSII2\_160759). H.A.S and R.M.G. acknowledge partial support from NASA grants NNX10AD68G and NNX12AI55G. The authors thank the referee and Kathleen Kraemer for careful reading of the manuscript and valuable comments that helped improve the paper.This publication makes use of data products from the Two Micron All Sky Survey, which is a joint project of the University of Massachusetts and the Infrared Processing and Analysis Center/California Institute of Technology, funded by the National Aeronautics and Space Administration and the National Science Foundation and NASAs Astrophysics Data System Bibliographic Services. 

{\it Facilities: }\facility{2MASS ($JHK_S$)}, \facility{Spitzer} (IRAC, MIPS), \facility{UKIRT ($JHK_S$)}

\begin{turnpage}
\begin{deluxetable}{lccccccccccccccc}
\tabletypesize{\scriptsize}
\tablecaption{Source List for W51\label{sourcetableW51}}
\tablewidth{0pt}
\tablehead{\colhead{Source Name\tablenotemark{a}} & \colhead{R.A.} & \colhead{Decl.} & \colhead{$J$} & \colhead{$H$} & \colhead{$K_{s}$} & \colhead{$[3.6]$} & \colhead{$[4.5]$} & \colhead{$[5.8]$} & \colhead{$[8.0]$} & \colhead{$[24]$} & \colhead{} & \colhead{} & \colhead{SED} & \colhead{} & \colhead{R-}\\
\colhead{} & \colhead{J2000.0} & \colhead{J2000.0} & \colhead{(mag)} & \colhead{(mag)} & \colhead{(mag)} & \colhead{(mag)} & \colhead{(mag)} & \colhead{(mag)} & \colhead{(mag)} & \colhead{(mag)} & \colhead{Type\tablenotemark{b}} & \colhead{Method\tablenotemark{c}} & \colhead{slope $\alpha$} & \colhead{Sigma\tablenotemark{d}} & \colhead{Squared}\\
\colhead{} & \colhead{($\degr$)} & \colhead{($\degr$)} & \colhead{} & \colhead{} & \colhead{} & \colhead{} & \colhead{} & \colhead{} & \colhead{} & \colhead{} & \colhead{} & \colhead{} & \colhead{} & \colhead{} & \colhead{}}
\startdata
SSTOERC$\_$G049.3793$-$0.2560 & 290.75937 & 14.47431 & 17.58(03) & 15.17(02) & 13.79(02) & 11.76(20) & 11.28(17) & \nodata & \nodata & 0.62(17) & 0 & 1 & 2.02 & 0.36 & 0.99 \\
SSTOERC$\_$G048.8881$-$0.2638 & 290.52740 & 14.03759 & 17.28(03) & 15.13(02) & 13.90(02) & 11.74(02) & 11.09(02) & 6.11.26(18) & \nodata & \nodata & 1 & 1 & 0.81 & 0.29 & 0.84 \\
SSTOERC$\_$G048.6453$-$0.1257 & 290.28412 & 13.88838 & 15.61(02) & 14.09(02) & 13.30(02) & 12.80(03) & 12.44(04) & \nodata &  \nodata &  \nodata & 2 & 1 & -1.75 & 0.06 & 1.00 \\
SSTOERC$\_$G048.6287+0.2361 & 289.94690 & 14.04359 & 14.08(02) & 12.27(03) & 11.38(02) & 10.79(01) & 10.60(01) & 10.47(07) & 9.87(18) & 4.50(16) & 3 & 1 & -1.71 & 0.30 & 0.90 \\
SSTOERC$\_$G049.1638$-$0.0498 & 290.46670 & 14.38155 & \nodata & 18.25(10) & 16.99(08) & 13.70(12) & 13.16(09) & \nodata & \nodata & \nodata & 4 & 2 & \nodata & \nodata & \nodata \\
SSTOERC$\_$G049.0780+0.5502 & 289.87799 & 14.58771 & 17.23(03) & 16.03(03) & 14.75(03) & 14.01(09) & 13.70(14) & \nodata & \nodata & \nodata & 5 & 2 & \nodata & \nodata & \nodata \\
SSTOERC$\_$G049.1458$-$0.0561 & 290.46362 & 14.36270 & 14.66(02) & 13.97(02) & 13.69(02) & 13.29(01) & 12.97(14) & \nodata & \nodata & 2.11(10) & 6 & 2 & \nodata & \nodata & \nodata \\  
SSTOERC$\_$G048.9561+0.4943 & 289.87006 & 14.45382 & 17.38(03) & 15.56(02) & 14.74(02) & 13.86(10) & 13.92(14) & \nodata & \nodata & \nodata & 99 & 1 & -1.48 & 0.33 & 0.99 \\
SSTOERC$\_$G048.8942$-$0.4828 & 290.72949 & 13.93967 & 14.65(02) & 13.68(02) & 13.19(02) & 12.40(03) & 12.37(04) & 11.70(09) & 10.29(07) & \nodata & 19 & 1 & \nodata & \nodata & \nodata \\
SSTOERC$\_$G049.3857$-$0.0146 & 290.5426 & 14.5938 & 17.89(04) & 16.65(03) & 15.53(03) & 13.30(04) & 12.18(04) & 11.77(09) & 10.99(15) & \nodata & 9 & 1 & \nodata  & \nodata & \nodata \\
SSTOERC$\_$G049.1445+0.2417 & 290.19168 & 14.5016 & 13.43(03) & 10.24(03) & 8.41(03) & 6.53(00) & 5.82(00) & 4.68(00) & 4.08(00) & \nodata & 20 & 1 & \nodata  &\nodata & \nodata\\
SSTOERC$\_$G049.3055+0.6570 & 289.89069 & 14.83866 & 10.18(03) & 8.48(05) & 7.40(03) & 6.68(01) & 6.18(00) & 5.59(00) & 5.30(00) & \nodata & 12 & 2 & \nodata &\nodata & \nodata \\
\enddata
\tablecomments{Table 13 is published in the electronic edition of the {\it Astrophysical Journal}. A portion is shown here for guidance regarding its form and content. Values in parentheses signify error in last 2 digits of magnitude value. Right ascension and Declination coordinates are J2000.0.} 
\tablenotetext{a}{Sources are named according to their Galactic longitude and latitude with the prefix SSTOERC, referring to Spitzer Space Telescope, Origin and Evolution of Rich Clusters project.}
\tablenotetext{b}{0; deeply embedded, 1; Class I, 2; Class II, 3; transition disks, 4; faint Class I, 5; faint Class II, 6; faint transition disks, 99; Class III and photospheres, 9; shocked gas emission, 29; AGNs, 19; PAH galaxies, 20; PAH-dominated sources, 12; AGBs, -100; unclassified.}
\tablenotetext{c}{Sources are flagged according to the method by which they were classified.1; \citet{gut09} classification, 2; AGBs, faint YSO elimination method (see Section 2.3).}
\tablenotetext{d}{ Sigma values signify the error in last two digits of the SED slope value.}
\end{deluxetable}
\end{turnpage}
\clearpage

\begin{turnpage}
\begin{deluxetable}{lccccccccccccccc}
\tabletypesize{\scriptsize}
\tablecaption{Source List for W43\label{sourcetableW43}}
\tablewidth{0pt}
\tablehead{\colhead{Source Name\tablenotemark{a}} & \colhead{R.A.} & \colhead{Decl.} & \colhead{$J$} & \colhead{$H$} & \colhead{$K_{s}$} & \colhead{$[3.6]$} & \colhead{$[4.5]$} & \colhead{$[5.8]$} & \colhead{$[8.0]$} & \colhead{$[24]$} & \colhead{} & \colhead{} & \colhead{SED} & \colhead{} & \colhead{R-}\\
\colhead{} & \colhead{J2000.0} & \colhead{J2000.0} & \colhead{(mag)} & \colhead{(mag)} & \colhead{(mag)} & \colhead{(mag)} & \colhead{(mag)} & \colhead{(mag)} & \colhead{(mag)} & \colhead{(mag)} & \colhead{Type\tablenotemark{b}} & \colhead{Method\tablenotemark{c}} & \colhead{slope $\alpha$} & \colhead{Sigma\tablenotemark{d}} & \colhead{Squared} \\
\colhead{} & \colhead{($\degr$)} & \colhead{($\degr$)} & \colhead{} & \colhead{} & \colhead{} & \colhead{} & \colhead{} & \colhead{} & \colhead{} & \colhead{} & \colhead{} & \colhead{} & \colhead{} & \colhead{} & \colhead{}}
\startdata
SSTOERC$\_$G031.2284$-$0.0191& 282.09952 & -1.52473 & 19.87(25) & 18.72(21) & 17.84(23) & \nodata & 12.25(07) & 10.63(11) & 9.60(14) & 4.58(08) & 0 & 1 & 1.41 & 0.32 & 0.90 \\
SSTOERC$\_$G030.4270+0.4739 & 281.29474 & -2.01290 & 16.28(11) & 18.09(09) & 15.07(03) & 12.87(08) & 12.48(09) & \nodata & \nodata & 5.89(24) & 1 & 1 & 0.72 & 0.23 & 0.98 \\
SSTOERC$\_$G029.5402+0.4199 & 280.93744 & -2.82646 & 17.41(03) & 15.09(02) & 13.92(02) & 12.60(04) & 11.94(03) & 11.29(07) & 10.36(08) & 7.58(12) & 2 & 1 & -0.35 & 0.04 & 0.98 \\
SSTOERC$\_$G030.3208+0.9046 & 280.86285 & -1.91069 & \nodata & \nodata & \nodata & 10.35(01) & 10.29(01) & 10.25(03) & 10.01(05) & 7.49(14) & 3 & 1 & -2.41 & 0.27 & 1.00 \\
SSTOERC$\_$G030.3622$-$0.2990 & 281.95340 & -2.42323 & \nodata & 18.28(08) & 16.53(05) & 13.81(09) & 12.78(03) & \nodata & \nodata & 6.28(28) & 4 & 2 & \nodata & \nodata & \nodata  \\
SSTOERC$\_$G029.6605+0.1712 & 281.21396 & -2.83302 & 17.38(03) & 15.81(02) & 14.66(02) & 13.35(08) & 12.81(07) & \nodata & \nodata & 7.75(19) & 5 & 2 & \nodata & \nodata & \nodata \\
SSTOERC$\_$G029.4575$-$0.1408 & 281.39908 & -3.15610 & 14.97(02) & 14.00(02) & 13.56(02) & 13.20(09) & 13.08(09) & \nodata & \nodata & 7.59(12) & 6 & 2 & \nodata & \nodata & \nodata \\ 
SSTOERC$\_$G030.5678+0.8209 & 281.05017 & -1.72920 & 12.72(03) & 10.72(03) & 9.89(02) & 9.32(01) & 9.45(01) & 9.21(02) & 9.17(03) & 9.63(03) & 99 & 1 & -2.62 & 0.28 & 0.99 \\
SSTOERC$\_$G030.6965+0.7979 & 281.12946 & -1.62522 & 14.32(02) & 13.39(02) & 12.96(02) & 12.39(03) & 12.40(03) & 11.68(10) & 10.19(07) & 7.56(08) & 19 & 1 & \nodata & \nodata & \nodata \\
SSTOERC$\_$G030.7902+0.2053 & 281.69977 & -1.81238 & \nodata & 18.53(15) & 15.41(03) & 11.39(02) & 10.05(02) & 9.85(05) & 9.61(09) & \nodata & 9 & 1 & \nodata  & \nodata & \nodata \\
SSTOERC$\_$G030.2314+0.8732 & 280.84994 & -2.00452 & \nodata & \nodata & \nodata & 8.89(01) & 9.21(01) & 8.84(01) & 8.77(02) & 9.33(02) & 20 & 1 & \nodata  &\nodata & \nodata\\
SSTOERC$\_$G030.7677+0.4362 & 281.48395 & -1.72702 & \nodata & 13.36(05) & 10.94(04) & 8.85(01) & 8.58(01) & 8.25(01) & 7.82(01) & 8.40(27) & 12 & 1 & \nodata &\nodata & \nodata\\
\enddata
\tablecomments{Table 14 is published in the electronic edition of the {\it Astrophysical Journal}. A portion is shown here for guidance regarding its form and content. Values in parentheses signify error in last 2 digits of magnitude value. Right ascension and Declination coordinates are J2000.0.} 
\tablenotetext{a}{Sources are named according to their Galactic longitude and latitude with the prefix SSTOERC, referring to Spitzer Space Telescope, Origin and Evolution of Rich Clusters project.}
\tablenotetext{b}{0; deeply embedded, 1; Class I, 2; Class II, 3; transition disks, 4; faint Class I, 5; faint Class II, 6; faint transition disks, 99; Class III and photospheres, 9; shocked gas emission, 29; AGNs, 19; PAH galaxies, 20; PAH-dominated sources, 12; AGBs, -100; unclassified.}
\tablenotetext{c}{Sources are flagged according to the method by which they were classified.1; \citet{gut09} classification, 2; AGBs, faint YSO elimination method (see Section 2.3).}
\tablenotetext{d}{ Sigma values signify the error in last two digits of the SED slope value.}
\end{deluxetable}
\end{turnpage}

\end{document}